\documentclass[12pt,aps,a4paper, floats,nofootinbib,amssymb,superscriptaddress]{revtex4}

\usepackage[sort&compress]{natbib}
\usepackage{epsf,epsfig}
\usepackage{amsmath}
\usepackage{hyperref}\usepackage{subfigure}
\usepackage{graphicx}
\usepackage{color}
\usepackage{mathrsfs}
\usepackage{enumitem}

\newcommand{\be}{\begin{equation}}
\newcommand{\ee}{\end{equation}}
\newcommand{\bea}{\begin{eqnarray}}
\newcommand{\eea}{\end{eqnarray}}

\newcommand{\ba}{\begin{array}}
\newcommand{\ea}{\end{array}}
\newcommand{\bi}{\begin{itemize}}
\newcommand{\ei}{\end{itemize}}

\newcommand{\redcross}{\color{red}-}
\newcommand{\doublepole}{{\color{red}{\bigotimes}}}
\newcommand{\doublerow}[2]{\hspace{0.3cm} \begin{matrix} #1 \\ #2 \end{matrix} \hspace{0.3cm}}

\newcommand{\myspace}{\vspace{0.2cm} \\}

\newcommand{\ih}{-\frac{i}{2}}

\makeatletter

\begin{document}


\title{\vspace*{1.in}
 Introducing tools to test Higgs interactions via $WW$ scattering. II.
  The coupled channel formalism and scalar resonances
\vspace*{0.5cm}
}

\author{I\~nigo  Asi\'ain}\email{iasiain@icc.ub.edu}\affiliation{Departament de F\'isica Qu\`antica i Astrof\'isica\,,
Institut de Ci\`encies del Cosmos (ICCUB), \\
Universitat de Barcelona, Mart\'i Franqu\`es 1, 08028 Barcelona, Spain}
\author{Dom\`enec Espriu}\email{espriu@icc.ub.edu}
\affiliation{Departament de F\'isica Qu\`antica i Astrof\'isica\,,
Institut de Ci\`encies del Cosmos (ICCUB), \\
Universitat de Barcelona, Mart\'i Franqu\`es 1, 08028 Barcelona, Spain}
\author{Federico Mescia}\email{mescia@ub.edu}\affiliation{Departament de F\'isica Qu\`antica i Astrof\'isica\,,
Institut de Ci\`encies del Cosmos (ICCUB), \\
Universitat de Barcelona, Mart\'i Franqu\`es 1, 08028 Barcelona, Spain}

\thispagestyle{empty}

\begin{abstract}
\vspace*{0.7cm}
\noindent
In this work we explore in detail the presence of scalar resonances in the $WW$ fusion process in the context of the LHC experiments
working in the theoretical framework provided by Higgs effective field theories (HEFTs). While the phenomenology of vector resonances
is reasonably understood in the framework of Weinberg sum-rules and unitarization studies, scalar resonances are
a lot less constrained 
and, more importantly do depend on HEFT low-energy effective couplings different from the ones of vector resoances that are difficult to constrain
experimentally. More specifically, unitarization techniques combined with the requirement of causality allows us to set nontrivial bounds on Higgs self-interactions. This is due to the need for considering coupled channels in the
scalar case along the unitarization process.
As a byproduct, we can gain some relevant information on the Higgs sector from $WW\to WW$ elastic  processes
without needing to consider two-Higgs production. 

\end{abstract}

\maketitle

\newpage
\thispagestyle{empty}
\tableofcontents

\thispagestyle{empty}
\newpage
 \setcounter{page}{0}


\section{Introduction}
\label{sec:intro}
In a companion paper Ref. \cite{Asiain:2021lch}, we highlighted the importance of the $W_LW_L$ scattering in investigating
the Higgs self-couplings, and therefore the Higgs potential, at the LHC in the framework of the Higgs effective field theory (HEFT).
We argued that unitarization of the amplitudes
was a convenient ---sometimes even necessary--- ingredient in this analysis. Summarizing, there are two reasons for this. On the one hand,
departures from the minimal Standard Model (SM) typically lead to violations of unitarity at large energies with fast rising
amplitudes.
Taking into account that the fundamental theory has presumably to be renormalizable and unitary, this may lead to hypersensitivity
to deviations of the effective theory coefficients with respect to their SM values. Even if this is not the case (for instance, because deviations with respect to the SM are tiny), it is known that lack of unitarity typically generates resonances in various
channels, which in a sense is the way the effective theory has to remember that it derives from a {\it bona fide} microscopic theory.
The properties of these resonances are typically very sensitive even to some small deviations with respect the SM and thus worth
investigating. \myspace
In Ref.~\cite{Asiain:2021lch}, we listed and renormalized all the suitable on-shell local operators of the vector and scalar
sector of the HEFT describing at low energies an extended electroweak symmetry breaking sector contributing to $2\to 2$
processes. Similar results were also reported in Ref.~\cite{Herrero:2021iqt} in the off-shell case, but without consideration
of the Higgs self-coupling. In Ref.~\citep{Herrero:2022krh} the complete process $W_L^{+}W_L^{-}\to hh$ where the triple Higgs 
coupling contributes at tree level and its renormalization plays a role, was in turn considered confirming the results 
in Ref.~\cite{Asiain:2021lch}. We also studied the presence of vector resonances ($IJ=11$) in the spectrum and their characteristics.
These resonances appeared after unitarization of the $WW$ elastic partial waves that would otherwise grow uncontrolled with
the centre-of-mass energy within the HEFT framework.\myspace
Unitarization of the amplitudes was carried out, making use of the Inverse Amplitude Method
(IAM) ~\cite{Truong:1988zp, Dobado:1989qm, Dobado:1996ps, Oller:1997ng, Guerrero:1998ei, Oller:1998hw,
  Dobado:2001rv, Corbett:2015lfa, Garcia-Garcia:2019oig, Salas-Bernardez:2020hua}, where the appearance of resonances can be understood
after the resummation of an infinite chain of bubble diagrams, hence dynamically. In the vector case, the only such
resummation possible is with $I=1$
intermediate states, $WW\to ZZ\to WW\to\ldots \to ZZ$, but for the scalar amplitude, $I=0$ double-Higgs state
insertions are permitted, leading to a chains of bubbles of the form $WW\to hh\to ZZ\to\ldots \to ZZ$. The details of
how to build the multichannel version of the IAM will be specified in the forthcoming sections, and the interested reader may find more
information in Refs.~\cite{Oller:1998hw, Oller:2000ma}.\myspace
It should be clear that unitarization in the $IJ=00$ channel and the ensuing possible resonances are a very promising tool to study and
eventually set relevant constraints on the Higgs self-couplings and all other parameters in an effective theory. As stated, a given
set of parameters of the effective theory typically leads to the appearance of resonances, required by unitarity.
If these resonances happen to have a low mass and 
should have already been experimentally seen, the absence of detection should translate into bounds on the effective parameters.
On the other hand, it may happen (and it does happen) that a given effective theory gives rise to unphysical resonances, located
in the first Riemann sheet. The corresponding set of parameters can also be excluded as no fundamental microscopic theory should
give rise to acausal behavior.
Thus, unitarization and resonances are important not only to reconcile effective theories with experiment but also to set 
bounds and exclude regions of parameter space. This is one of the purposes of the present work. \myspace
When compared to works that made use of the Equivalence
Theorem \cite{He:1993qa,  Grosse-Knetter:1994lkr, Dobado:1994vr, Dobado:1993dg, Chanowitz:1985hj, Gounaris:1986cr,
  Lee:1977eg, Cornwall:1974km} (ET) in its extreme version, where no transverse modes of the electroweak gauge bosons were allowed
inside the loops \cite{Espriu:2012ih,Delgado:2013hxa}, it was found that including the complete $\mathcal{O}(g)$ calculation 
did not noticeably modify masses and widths of the vector resonances. Consequently, we found a clear hierarchy among the HEFT coefficients as far as the properties of the resonances is concerned:
the positions of the vector poles are mainly controlled by
the parameters surviving in the extreme $g=0$ ET limit, namely, $a_4$ and $a_5$, with only small variations
when the new $\mathcal{O}(g)$ operators are introduced via $a_3$ and $\zeta$ parameters \cite{Asiain:2021lch} (see below for
the proper definition
of these parameters). This is important because the number of free parameters explodes in effective theories and it is relevant to know beforehand
those that may be more relevant for phenomenology.\myspace
In this work, we begin by briefly reviewing the basic setup and notation of the HEFT. Then, we proceed to examining the existing bounds
on the various parameters involved in the HEFT, paying special attention to those involving the Higgs. In Sec.~\ref{sec:processes}
the isospin projections and several technical aspects of the calculations are reviewed in a cursory manner (including several comments
on the approximations made). Section~\ref{sec:unitarization} is devoted to a discussion, in the present context, on the IAM when
several coupled channels are present. A comparison and a discussion on how the presence of coupled channels alters previously
existing results is included there, too. The next section is devoted to a in-depth analysis of the various resonances that appear
for a range of parameters.
It is seen that for certain values of the Higgs potential causality violations appears. This implies that there cannot be a consistent
microscopic, i.e., fundamental, theory whose low-energy realization ---the HEFT--- is described by such values.
Other sets of parameters can be excluded because they would lead to resonances that should have been observed.

\section{Effective Lagrangian}\label{sec:lagrangian}
The HEFT is a nonlinear gauge effective field theory (EFT) that includes a light (with respect to the
scale of new physics) Higgs-like field, and it is a natural extension of the Electroweak Chiral Lagrangian \cite{Dobado:2019fxe, Contino:2013kra,Giudice:2007fh, Dobado:1990zh, Longhitano:1980iz}.
All the details about how to build this effective theory that only draws from the local and global properties of the electroweak
sector can be found in the references listed in Ref. \cite{Asiain:2021lch}. We will work under the approximation that
the custodial symmetry remains exact and consequently the electromagnetism is removed from our theory along
with other possible sources of custodial breaking.
In practice, it will suffice to set $g^{\prime}=0$ and leave custodial breaking operators aside.\myspace
The low-energy degrees of freedom of this EFT are the electroweak gauge bosons $W^{\pm},Z$; their associated Goldstone bosons arising
from the spontaneous symmetry breaking $\omega^{\pm},z$; and the light Higgs, which in the nonlinear realization of the chiral symmetry
remains a singlet in clear contrast to the $SU(2)_L$ doublet of the linear case. The theory is built as an expansion of powers of
the momenta of the external Goldstones that quickly leads to a violation of unitarity of the amplitudes even within the range of
convergence of the EFT.\myspace
The relevant pieces for the calculations of on-shell $2\to 2$ processes in the scalar and vector sector of the HEFT up to next-to-leading (NLO) and under
the assumptions mentioned before are listed hereunder
\begin{equation}\label{eq: lag2}
\begin{split}
  \mathcal{L}_2 =&-\frac{1}{2g^2}\text{Tr}\left(\hat{W}_{\mu\nu}\hat{W}^{\mu\nu}\right)-
\frac{1}{2g^{\prime 2}}\text{Tr}\left(\hat{B}_{\mu\nu}\hat{B}^{\mu\nu}\right)
  +\frac{v^2}{4}\mathcal{F}(h)\text{Tr}\left(D^{\mu}U^{\dagger}D_{\mu}U\right)\\
  &+\frac{1}{2}\partial_{\mu}h\partial^{\mu}h -V(h) \vspace{0.2cm} \\
\end{split}
\end{equation}
\begin{equation}\label{eq: lag4}
\begin{split}
  \mathcal{L}_4 =&-i a_3\text{Tr}\left(\hat{W}_{\mu\nu}\left[V^{\mu},V^{\nu}\right]\right)
  +a_4 \left(\text{Tr}\left(V_{\mu}V_{\nu}\right)\right)^2
  +a_5 \left(\text{Tr}\left(V_{\mu}V^{\mu}\right)\right)^2+\frac{\gamma}{v^4}\left(\partial_{\mu}h\partial^{\mu}h\right)^2\\
  &+\frac{\delta}{v^2}\left(\partial_{\mu}h\partial^{\mu}h\right)\text{Tr}\left(D_{\mu}U^{\dagger}D^{\mu}U\right)
  +\frac{\eta}{v^2}\left(\partial_{\mu}h\partial_{\nu}h\right)\text{Tr}\left(D^{\mu}U^{\dagger}D^{\nu}U\right)\\
&+i\frac{\zeta}{v}\,\text{Tr}\left(\hat{W}_{\mu\nu}V^{\mu}\right)\partial^{\nu}h
\end{split}
\end{equation}
with the building blocks
\begin{equation}\label{eq: building_blocks}
\begin{split}
  &U=\exp\left(\frac{i\omega^a\sigma^a}{v}\right) \in SU(2)_V, \quad V_{\mu}=D_{\mu}U^{\dagger}U,
  \quad \mathcal{F}(h)=1+2a\left(\frac{h}{v}\right)+b\left(\frac{h}{v}\right)^2+ \ldots , \\
  &D_{\mu}U=\partial_{\mu}U+i \hat{W}_{\mu} U, \quad \hat{W}_{\mu}=g\frac{\vec{W}_{\mu}\cdot\vec{\sigma}}{2},
  \quad \hat{W}_{\mu\nu}=\partial_{\mu}\hat{W}_{\nu}-\partial_{\nu}\hat{W}_{\mu}+i\left[\hat{W}_{\mu},\hat{W}_{\nu}\right], \\
  &V(h)=\frac{1}{2}M_h^2h^2+\lambda_3 v h^3+\frac{\lambda_4}{4}h^4+ \ldots
\end{split}
\end{equation}
The so-called \textit{anomalous}, i.e., not present in the SM, chiral couplings follow the notation in Ref. \citep{Asiain:2021lch}.
The $\omega^a$ fields are the Goldstone degrees of freedom that are gathered in the unitary $U$ matrix included in the custodial
group $SU(2)_L\times SU(2)_R/SU(2)_V$. \myspace
We will use the parametrization $\lambda_{3,4}=d_{3,4}\lambda_{SM}$, where $\lambda_{SM}$ is the SM Higgs self-interaction that defines
its mass $M_h^2=2\lambda_{SM} v^2$. In our beyond-SM (BSM) theory, the departures from the SM Higgs potential are carried out by
the dimensionless couplings $d_3$ and $d_4$. These last parameters will play a key role in the present study since they enter
at tree level in the two-Higgs production unlike in the $I=1$ case.\myspace
In our previous work \cite{Asiain:2021lch}, the complete list of counterterms associated to the electroweak and chiral parameters
of the custodial-preserving Lagrangian was presented up to order $s^2$. In the Landau gauge and by making use of the
Equivalence Theorem, we performed the one-loop on-shell renormalization of the three processes involved. We cross-checked our results
with previous work that assumed a massless scenario \cite{Delgado:2013hxa} and also with results in an off-shell
basis \cite{Gavela:2014uta,Herrero:2021iqt}. \myspace
Finally,
\begin{equation}\label{eq: complete_lag}
\mathcal{L}=\mathcal{L}_2+\mathcal{L}_4+\mathcal{L}_{GF}+\mathcal{L}_{FP}
\end{equation}
where the last two pieces correspond to the gauge fixing and Faddeev-Popov terms, respectively, essential for adding the
quantum corrections. Regarding this matter, we will be working in the Landau gauge $(\xi=0)$ with massless Goldstones and no
interactions
among the Goldstone sector and the Faddeev-Popov ghosts since they are proportional to the gauge parameter. There is no
fundamental reason to do so, but this approach makes the calculations simpler. In the HEFT, there are no interactions
between the Higgs and the ghost sector either, the former being a singlet in the nonlinear realization of the chiral symmetry
and thereupon absent in both the gauge fixing and Faddeev-Popov terms.\myspace
\section{Relevant amplitudes}\label{sec:processes}
Let us begin by reviewing the experimental situation regarding the existing bounds on the couplings of the HEFT. Not all low-energy
constants are constrained. At present, there are no bounds available on various $\mathcal{O}(p^4)$ parameters. It is important
to note that the Higgs potential parameters $\lambda_3$ and $\lambda_4$ (i.e., $d_3$ and $d_4$) are poorly constrained and not constrained at all,
respectively, from an experimental point of view. See the Table \ref{table: exp} for existing information of the parameters of the
HEFT.
\begin{table}[tb]  
\begin{center}
\renewcommand{\arraystretch}{1.2}
\begin{tabular}{|c|c|c| }
\hline
Couplings & Ref. & Experiments\\
\hline \hline
$0.89<a<1.13$  & \cite{deBlas:2018tjm}
& LHC  \\ \hline
$-0.76<b<2.56$ & \cite{ATLAS:2020jgy} & ATLAS \\ \hline
$-3.3<d_3<8.5$ &\cite{CMS:2020tkr} & CMS \\ \hline
$d_4$ & - & - \\ \hline
 $|a_1|<0.004$  &
 \cite{Tanabashi:2018oca} & LEP ($S$ parameter)\\ \hline
  $-0.06<a_2-a_3<0.20$&\cite{Almeida:2018cld} & LEP \& LHC  \\ \hline
 $-0.0061<a_4 < 0.0063$&\cite{CMS:2019uys} & CMS (from $WZ\to 4l$) \\ \hline
$|a_5|<0.0008$ &  \cite{Sirunyan:2019der} & CMS (from $WZ/WW\to 2l2j$) \\  \hline
\end{tabular}
\caption{{\small
    Current experimental constraints on bosonic HEFT anomalous
    couplings at 95\% C.L. See our work \citep{Asiain:2021lch} about the issue to extract the $a_4$ bound from the CMS
    analysis of Ref. \cite{Sirunyan:2019der}. Note that $d_4$ is not constrained at all at present from the experimental
    point of view.}} \label{table: exp}
\end{center}
\end{table}
However, it should be mentioned that from the results of Ref. \cite{Asiain:2021lch} it becomes clear that some of the parameters in the
$\mathcal{O}(p^4)$ HEFT are not very relevant in determining the properties of the dynamical resonances that appear and consequently
they only play a marginal role in the restoration of unitarity. Indeed, of all the effective couplings contributing to the
$I=1$ channel, $a_3$ and particularly $\zeta$ are not that important in determining the mass and width of the resonances. In the
$I=0$ case studied here, various other couplings enter, and it is {\em per se} interesting to assess the relevance of each of them, as they are in principle unknown except for general order-of-magnitude estimates. Needless to
say, it is particularly interesting to assess the relevance of the Higgs scalar couplings for the reasons already mentioned. \myspace
In the $I=0$ channel, there are three $2\to 2$ processes that need to be taken into account:  $WW\to WW$, where by $WW$ we refer
generically to any initial state with two $W$ or two $Z$ that is compatible with the prescribed isospin projection (we refer to
this process as the
elastic one); $WW \to hh$; and $hh \to hh$. Along the unitarization process, all three become coupled and need to
be considered.\myspace
The first step is to consider the tree plus one-loop perturbative contribution to the three subprocesses, including the
necessary counterterms \citep{Asiain:2021lch}. The relevant amplitudes are described below.
\subsection{Tree-level amplitudes and counterterms}\label{sec:amplitudes}
In this section we show the tree-level amplitudes of the processes relevant for the study. The level of precision that we aim for requires
having physical gauge bosons in the external states, as said before in the Introduction. For reasons that will be clear immediately below,
we just need to compute the three processes presented here to get the suitable scalar-isospin projection. Because of the lengthy expressions
that we get, we give the results split in the different channels participating in the process using the following notation that
was used in Ref. \citep{Asiain:2021lch}: a superindex indicates the different processes labeled
as $WW$ for $W^+W^-\to ZZ$, $Wh$ for $W^+W^-\to hh$, and $hh$ for $hh\to hh$. Also, each amplitude carries a
subindex ${xy}$ that represents a process with a particle $y$ propagating in the $x$ channel.
In the case with $x=c$ and no $y$, $\mathcal{A}_{c}$ represents the contact interaction of the four external particles.
For instance, the amplitude $\mathcal{A}^{WW}_{sh}$ represents a Higgs exchanged in the $s$ channel of $W^+W^-\to ZZ$ scattering.\\
The tree-level amplitudes corresponding to the processes under discussion are as follows:
\subsubsection{$W^{+}W^{-}\to ZZ$}
\begin{equation}\label{eq: tree_WW}
\begin{split}
\mathcal{A}^{WW}_c=&g^2\left(\left((-2 a_3+a_4)g^2+1\right)\left(\left(\varepsilon_1\varepsilon_4\right) \left(\varepsilon_2\varepsilon_3\right)+ \left(\varepsilon_1 \varepsilon_3\right) \left(\varepsilon_2 \varepsilon_4\right)\right)\right.\\
&\left.+2 \left((2 a_3+a_5)g^2-1\right) (\varepsilon_1\varepsilon_2) \left(\varepsilon_3\varepsilon_4\right)\right)\\
\mathcal{A}^{WW}_{sh}=&-\frac{a^2g^2M_W^2\left(\varepsilon_1\varepsilon_2\right)\left(\varepsilon_3\varepsilon_4\right)}{(p_1+p_2)^2-M_H^2}+\frac{ag^4\zeta}{4((p_1+p_2)^2-M_H^2)}\left[2(\varepsilon_3\varepsilon_4)\left((p_1\varepsilon_2)(p_2\varepsilon_1) \right.\right.\\
&\left.\left. -(\varepsilon_1\varepsilon_2)(p_1+p_2)^2\right)+2(\varepsilon_1\varepsilon_2)(p_3\varepsilon_4)(p_4\varepsilon_3)\right]\\
\mathcal{A}^{WW}_{tW}=&-\frac{(1-2a_3g^2)g^2}{(p_1-p_3)^2-M_W^2} \left[-4\left((\varepsilon_1\varepsilon_2)(p_1\varepsilon_3)(p_2\varepsilon_4)+(\varepsilon_1\varepsilon_4)(p_1\varepsilon_3)(p_4\varepsilon_2)\right.\right.\\
&\left. +(\varepsilon_2\varepsilon_3)(p_3\varepsilon_1)(p_2\varepsilon_4)+(\varepsilon_3\varepsilon_4)(p_3\varepsilon_1)(p_4\varepsilon_2) \right)\\&+2\left( (\varepsilon_2\varepsilon_4)\left( (p_1\varepsilon_3)(p_2+p_4)\varepsilon_1+(p_3\varepsilon_1)(p_2+p_4)\varepsilon_3\right)\right.\\
&\left. +(\varepsilon_1\varepsilon_3)((p_2\varepsilon_4)(p_1+p_3)\varepsilon_2+(p_4\varepsilon_2)(p_1+p_3)\varepsilon_4)\right)\\
&\left.-(\varepsilon_1\varepsilon_3)(\varepsilon_2\varepsilon_4)((p_1+p_3)p_2+(p_2+p_4)p_1) \right]\\
\mathcal{A}^{WW}_{uW}=&\mathcal{A}_{tW}(p_3\leftrightarrow p_4, \varepsilon_3 \leftrightarrow \varepsilon_4)
\end{split}
\end{equation}
where $\varepsilon_i$ is the abbreviation for $\varepsilon_L(p_i)$.

\subsubsection{$WW\to hh$}

\begin{equation}\label{eq: tree_Wh}
\begin{split}
\mathcal{A}^{Wh}_c=&\frac{g^2\,b}{2}(\varepsilon_1\varepsilon_2)-\frac{g^2\,\eta}{v^2}((\varepsilon_1 p_4)(\varepsilon_2 p_3)+(p_3\varepsilon_1)(\varepsilon_2 p_4))-\frac{2g^2\,\delta}{v^2}(p_3 p_4)(\varepsilon_1\varepsilon_2)\\
&+\frac{g^2\,\zeta}{v^2}((\varepsilon_1\varepsilon_2)(p_1+p_2)^2-2(p_1\varepsilon_2)(p_2\varepsilon_1))\\
\mathcal{A}^{Wh}_{sh}=&\frac{3g^2M_h^2}{2((p_1+p_2)^2-M_h^2)}\left(a(\varepsilon_1\varepsilon_2)+\frac{\zeta}{v^2}((\varepsilon_1\varepsilon_2)(p_1+p_2)^2-2(p_1\varepsilon_2)(p_2\varepsilon_1))\right)\\
\mathcal{A}^{Wh}_{t\omega}=&\frac{2a^2g^2+a\zeta g^4}{2(p_1-p_3)^2}\left((p_3\varepsilon_1)(p_4\varepsilon_2)\right)\\
\mathcal{A}_{tW}=&\frac{a^2g^2M_W^2}{((p_1-p_3)^2-M_W^2)}\left(\varepsilon_1\varepsilon_2+\frac{(p_4\varepsilon_2)(\varepsilon_1 p_3)}{(p_1-p_3)^2}\right)+\frac{ag^4\zeta}{2((p_1-p_3)^2-M_W^2)}\left( 2M_h^2(\varepsilon_1\varepsilon_2)\right. \\
&\left. -(p_4\varepsilon_2)(p_2\varepsilon_1)-(\varepsilon_1p_3)(\varepsilon_2p_3)+M_W^2\frac{(p_4\varepsilon_2)(\varepsilon_1p_3)}{(p_1-p_3)^2}\right)\\
\mathcal{A}^{Wh}_{u\omega}=&\mathcal{A}^{Wh}_{t\omega}(p_3\leftrightarrow p_4)\\
\mathcal{A}^{Wh}_{uW}=&\mathcal{A}^{Wh}_{tW}(p_3\leftrightarrow p_4)\\
\end{split}
\end{equation}

\subsubsection{$hh\to hh$}

\begin{equation}\label{eq: tree_hh}
\begin{split}
\mathcal{A}^{hh}_c=&\frac{8\gamma}{v^4}\left((p_1 p_4)(p_2 p_3)+(p_1 p_3)(p_2 p_4)+(p_1 p_2)(p_3 p_4)\right)-6\lambda_4\\
\mathcal{A}^{hh}_{sh}=&-\frac{36\lambda_3^2v^2}{(p_1+p_2)^2-M_h^2}\\
\mathcal{A}^{hh}_{th}=&\mathcal{A}^{hh}_{sh}(p_2\leftrightarrow -p_3)\\
\mathcal{A}^{hh}_{uh}=&\mathcal{A}^{hh}_{sh}(p_2\leftrightarrow -p_4)
\end{split}
\end{equation}\myspace
In all the previous expressions, the various couplings and parameters $v, a, b, a_3, a_4, \ldots $ contain the corresponding
counterterms needed for the one-loop renormalization:
$v \to v+\delta v, a\to a+ \delta a, b\to b + \delta b, a_3 \to a_3 + \delta a_3, \ldots $ The set of counterterms required
to render finite the physical amplitudes is provided in the Appendix. \myspace
In Ref. \cite{Asiain:2021lch} the interested reader can find a more detailed construction of the
isospin projections (isoscalar, isovector, and isotensor) of the $2\rightarrow 2$ processes that concern us here. In this section, we
will only summarize the main points relevant for the $I=0$ case. We emphasize that in the scalar case, the two-Higgs final
state is also present and that through the coupled channel the unitarization mechanism contributes to the elastic $WW\to WW$ channel,
and this is a relevant fact because this turns out to be easier to handle experimentally than processes involving two-Higgses final
states, as we will see in the coming sections. This opens an interesting window: the dynamical resonances potentially present in
elastic $WW$ scattering carry information from Higgs production. We will exploit this potential below.\myspace
The generic process $W^a_LW^b_L\to W^c_LW^d_L$, assuming an exactly preserved custodial symmetry and using Bose symmetry,
can be written
\begin{equation}\label{eq: isos_general}
\mathcal{A}^{abcd}=\delta^{ab}\delta^{cd}\mathcal{A}\left(p^a,p^b,p^c,p^d\right)
+\delta^{ac}\delta^{bd}\mathcal{A}\left(p^a,-p^c,-p^b,p^d\right)+\delta^{ad}\delta^{bc}\mathcal{A}\left(p^a,-p^d,p^c,-p^b\right)
\end{equation}
which allows us to write the following amplitudes in the more familiar charge basis:
\begin{equation}\label{eq: isos_I1}
\begin{split}
\mathcal{A}^{+-00}&=\mathcal{A}(p^a,p^b,p^c,p^d)\\
\mathcal{A}^{+-+-}&=\mathcal{A}(p^a,p^b,p^c,p^d)+\mathcal{A}(p^a,-p^c,-p^b,p^d)\\
\mathcal{A}^{++++}&=\mathcal{A}(p^a,-p^c,-p^b,p^d)+\mathcal{A}(p^a,-p^d,p^c,-p^b).
\end{split}
\end{equation}
Given this, one can see that once the "fundamental" amplitude $W_L^+W_L^-\to Z_LZ_L$ has been computed, the other ones are
obtained simply by crossing symmetry.\myspace
Since we are interested in a partial-wave analysis of unitarity, it is suitable to build the fixed-isospin amplitudes of the process that read
\begin{equation}\label{eq: Ts}
\begin{split}
T_0&=3\mathcal{A}^{+-00}+\mathcal{A}^{++++}\\
T_1&=2\mathcal{A}^{+-+-}-2\mathcal{A}^{+-00}-\mathcal{A}^{++++}\\
T_2&=\mathcal{A}^{++++}
\end{split}
\end{equation}
In contrast to our previous work, we will now be interested in the scalar projection, $T_0$. \myspace
Within our theoretical framework, the Higgs is a weak isospin singlet $\left( I=0\right)$, so it is straightforward to write
the following relations for the $W^a_LW^b_L\to hh$ and $hh\to hh$ processes:
\begin{equation}\label{eq: isos_M0}
\mathcal{M}(W_L^a W_L^b \to hh)=\mathcal{M}^{ab}(p^a,p^b,p_{h,1},p_{h,2}),\qquad
T_{Wh,0}=\sqrt{3}\mathcal{M}^{+-},
\end{equation}
\begin{equation}\label{eq: isos_T0}
\mathcal{T}(hh\to hh)=\mathcal{T}(p_{h,1},p_{h,2},p_{h,3},p_{h,4})= \mathcal{T}_{hh,0},
\end{equation}
\subsection{One-loop real part: The Equivalence Theorem}
Taking into account that we will be eventually interested in exploring a large set of parameters, it is important to be equipped with
computational algorithms that run fast. This is one of the reasons that make convenient, in order to determine the real part of the
one-loop amplitude, to appeal to the so-called Equivalence Theorem. This allows us to replace the scattering of physical $W$'s by the
one of the corresponding Goldstone bosons. The details concerning this approximation can be found in Ref. \citep{Espriu:1994ep} and were also briefly reviewed in Ref. \citep{Asiain:2021lch}.\myspace
Another reason why it may be convenient to use the ET has to do with the convenience to make direct contact with some existing analytical
results that are only available in the ET limit and for $g=0$ where the expressions become analytically tractable, because a full one-loop calculation in the $g\neq 0$ limit involves expressions
that do not have a simple analytical continuation to the complex $s$ plane.\myspace
We chose to work in the 't Hooft-Landau gauge that is particularly simple from the renormalization point of view. As discussed in
Ref. \citep{Asiain:2021lch}, the results we obtain should be gauge invariant at the order where we are working, even if, in principle, the leading
order of the ET approximation is by itself not gauge invariant. Recall that the ET relies on the splitting of the polarization
vector $\epsilon^\mu_L = k^\mu/M_W + v^\mu$, with $v^\mu$ being of order $M_W/\sqrt{s}$. We note that we use the ET for the one-loop
correction only, not for the tree-level contribution that is calculated using physical $W$'s.
The one-loop correction to the partial wave is of $O(s^2)$ and
  the (potentially gauge-dependent) corrections to the ET might change the $O(s)$ contribution, but the latter ---tree level---
is calculated exactly without appealing to the ET. Therefore, gauge invariance is respected even if the splitting is itself not
gauge invariant. As a sanity check, where a comparison can be made, all counterterms agree with those computed in a general gauge. \myspace
A further check is provided by comparing  the imaginary part obtained in this way with the one computed exactly via the
optical theorem.\myspace
Even using the ET, the one-loop real part cannot be expressed when masses are not neglected in terms of simple functions and the results
are derived numerically. Fortunately, most of the effective coefficients enter only at order $s^2$, and then they are formally tree level
from a computational point of view, even though they give contributions of the same order as the one loop. This makes exploring the
parameter landscape simpler. An exception to this rule are the parameters $a$, $b$, $d_3$, and $d_4$ that enter at order $s$. Accordingly, every
modification of any of those requires a new one-loop calculation of the real part.\myspace
For further reference, we give below the expression for the tree and one-loop results in the limit $M_W=M_H=0$ \cite{Delgado:2013hxa}. They are
useful to identify physical poles in the full-fledged calculation that, as said, is not amenable to analytical continuation
to the appropriate Riemann sheet. In the limit where all the particles are massless and hence the SM values $g$ and $\lambda_{SM}$ are set to
zero, the amplitudes with $W$s in the external states vanish, and if we want to have some analytical expressions for the tree level for
our unitarity study, we are forced to go to the ET and place Goldstone bosons in the external legs.\myspace
The authors in Ref. \citep{Delgado:2013hxa} worked with the chiral parameters $\alpha$ and $\beta$, instead of $a$ and $b$, since
they introduced a vacuum-tilt extra free parameter, $\xi=\sqrt{\frac{v}{f}}$, interpolating between composite models ($v=f$) and
the SM limit ($f\to\infty$) where the new resonant states completely decouple from the theory (this vacuum-tilt parameter should not be
confused with the  gauge parameter). In fact, with the massless and naive custodial limit $g=g^{\prime}=0$ in Ref. \citep{Delgado:2013hxa} [also
known as \textit{naive Equivalence Theorem} (nET)], all
the gauge dependence disappears and all the amplitudes are trivially gauge invariant. This parametrization makes contact with ours with
the redefinitions $a=\alpha \sqrt{\xi}$ and $b=\beta\xi$. In our framework, only the electroweak scale is used for the Higgs mechanism
and weighs both Goldstone and Higgs fields, so we rewrite their amplitudes in the particular case: $\xi=1$, $a=\alpha$ and $b=\beta$.\myspace
The expressions are shown below and, in contrast to the full calculation previously described, due to the simplicity of the formulas,
we do not split the full amplitude in the different channels:

\subsubsection{$\omega\omega\to \omega\omega$ Massless limit:}
\begin{equation}\label{eq: A_tree}
\mathcal{A}^{tree}=\left(1-a^2\right)\frac{s}{v^2}+\frac{4}{v^4}\left(2a_5s^2+a_4(t^2+u^2)\right)
\end{equation}
\begin{equation}\label{eq: A_loop}
\mathcal{A}^{loop}=\frac{1}{576\pi^2 v^4}\left[f^W(s,t,u)s^2+(1-a^2)^2(g(s,t,u)t^2+g(s,u,t)u^2)\right]
\end{equation}
with the definitions
\begin{equation}
\begin{split}
f^W(s,t,u)&=20-40a^2+56a^4-72a^2b+36b^2+\Delta(12-24a^2+30a^4-36a^2b+18b^2)\\
&+(-18+36a^2-36a^4+36a^2b-18b^2)\log\left(\frac{-s}{\mu^2}\right)\\
&+3(1-a^2)^2\left[\log\left(\frac{-t}{\mu^2}\right)+\log\left(\frac{-u}{\mu^2}\right)\right]\\
\\
g(s,t,u)&=26+12\Delta-9\log\left(\frac{-t}{\mu^2}\right)-3\log\left(\frac{-u}{\mu^2}\right)
\end{split}
\end{equation}

\subsubsection{$\omega\omega\to hh$ Massless limit:}

\begin{equation}\label{eq: M_tree}
\mathcal{M}^{tree}=(a^2-b)\frac{s}{v^2}+\frac{2\delta}{v^4}s^2+\frac{\eta}{v^4}(t^2+u^2)
\end{equation}
\begin{equation}\label{eq: M_loop}
\mathcal{M}^{loop}=\frac{a^2-b}{576\pi^2v^2}\left[f^{WH}(s,t,u)\frac{s^2}{v^2}+\frac{a^2-b}{v^2}(g(s,t,u)t^2+g(s,u,t)u^2)\right]
\end{equation}
where
\begin{equation}
\begin{split}
f^{WH}(s,t,u)=&-8(-9+11a^2-2b)-6\Delta(-6+7a^2-b)-36(1-a^2)\log\left(\frac{-s}{\mu^2}\right)\\
&+3(a^2-b)\left(\log\left(\frac{-t}{\mu^2}\right)+\log\left(\frac{-u}{\mu^2}\right)\right)
\end{split}
\end{equation}
and the function $g(s,t,u)$ is the same as in the elastic case.

\subsubsection{$hh\to hh$ Massless limit:}
\begin{equation}\label{eq: T_tree}
\mathcal{T}^{tree}=\frac{2\gamma}{v^4}(s^2+t^2+u^2)
\end{equation}
Notice that this process has no $\mathcal{O}(p^2)$ contribution since in the massless Higgs limit, the triple self-coupling of
the Higgs vanishes and there is no diagram contributing to the process,
\begin{equation}
\mathcal{T}^{loop}=\frac{3(a^2-b)^2}{32\pi^2 v^4}\left(f^H(s)s^2+f^H(t)t^2+f^H(u)u^2\right)
\end{equation}
with
\begin{equation}
f^H(s)=2+\Delta-\log\left(\frac{-s}{\mu^2}\right)
\end{equation}

\subsection{One-loop imaginary part: The optical theorem}
\begin{figure}
\centering
\includegraphics[clip,width=13cm,height=9cm]{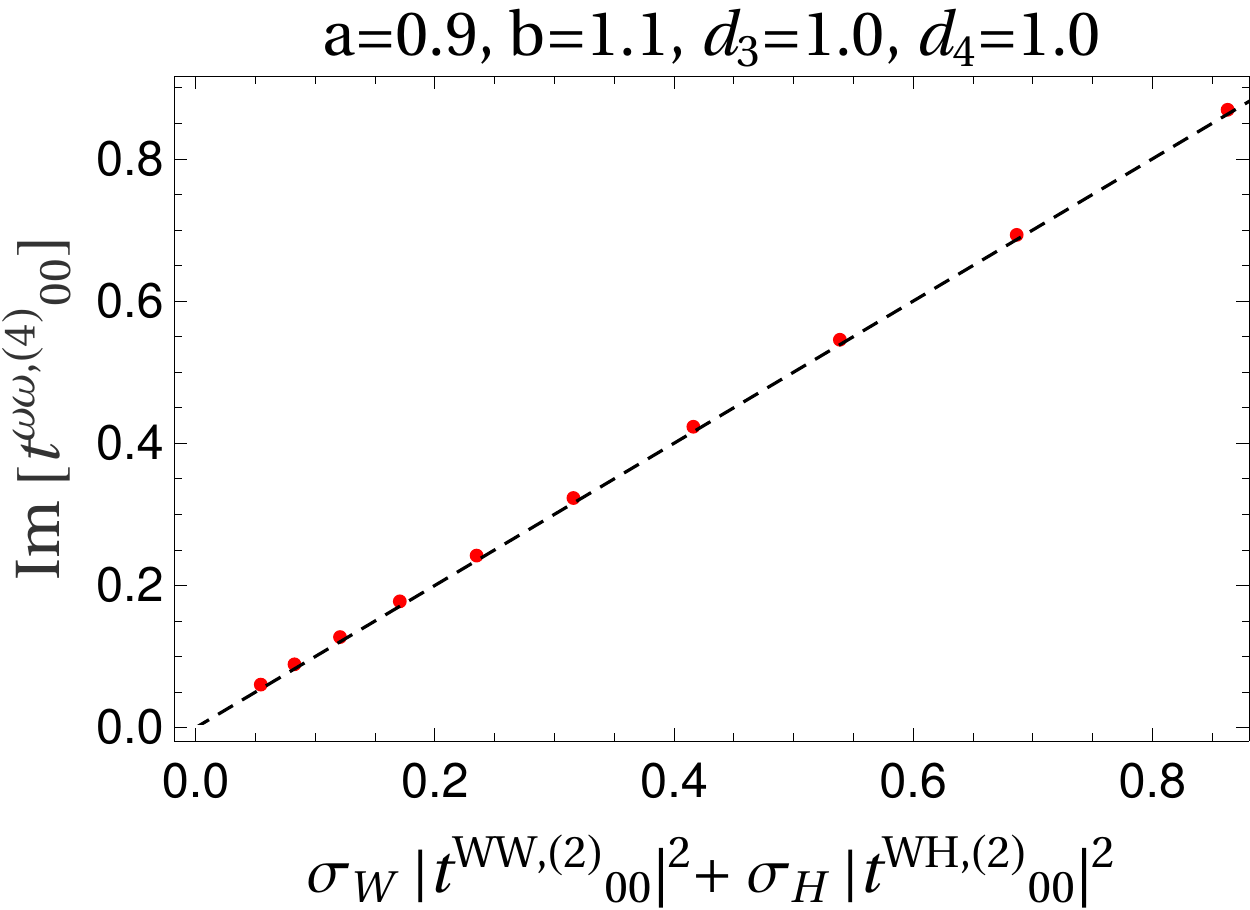}
\caption{\small{Figure showing perturbative unitarity (\ref{eq: perturbative_unitarity}) for the chiral couplings specified in the title.
    The data shown here are 10 red points equally spaced along the energy range $1500-3000$ GeV, within the validity region of the theory. The dashed line is the bisector of the first quadrant and the points that lie over it satisfy perturbative unitarity exactly.}}
\label{fig: perturbative_unitarity}
\end{figure}

With respect to the imaginary part of the one loop calculation, it is most easily determined exactly by using the optical theorem.
The details about our calculation of the imaginary part using this theorem are gathered in Ref. \citep{Asiain:2021lch} but for the scalar
case, we find one difference with respect to our previous study: an intermediate
double-Higgs $I=0$ state is now permitted.\myspace
Once we know the discontinuity of a complex amplitude, $\mathcal{A}(s)$, across the physical cut, we find
\begin{equation}\label{eq: OT}
Im\,\mathcal{A}(s)=\sum_{|\psi(I=0)>}\sigma (s)|\mathcal{A}(s)|^2
\end{equation}
where $\sigma (s)=\sqrt{1-\frac{(M_1+M_2)^2}{s}}$ is the two-body phase space. This allows us
to compute the imaginary part of any amplitude at the one-loop level from the tree-level result.\myspace
As an example, we show the different contribuitions in the full $I=0$ isospin amplitude in the process
$W_L^+W_L^-\to Z_LZ_L$:
\begin{equation}\label{eq: full_amp}
\mathcal{A}(W_L^+W_L^-\to Z_LZ_L)=\mathcal{A}_{tree}^{(2)}+\mathcal{A}_{tree}^{(4)}+\mathcal{A}_{loop}^{(4)}.
\end{equation}
$\mathcal{A}^{(2)}_{tree}+\mathcal{A}_{tree}^{(4)}$ is the full tree-level contribution (\ref{eq: tree_WW}) and $\mathcal{A}_{loop}^{(4)}$
is the one-loop amplitude
\begin{equation}
\mathcal{A}^{(4)}_{loop}=Re \left[\mathcal{A}^{(4)}_{loop}(\omega^+\omega^-\to zz)\right]+i \left(\sigma_W (s) |\mathcal{A}_{tree}^{(2)}|^2+
\sigma_H(s) |\mathcal{A}^{(2)}_{tree}(W_L^+W_L^-\to hh)|^2\right).
\end{equation}
$\mathcal{A}^{(2)}_{tree}(W_L^+W_L^-\to hh)$ in the imaginary part is the tree level amplitude of the VBS two-Higgs production,
equation (\ref{eq: tree_Wh}), and $\sigma_{W,H}=\sqrt{1-\frac{4M_{W,H}^2}{s}}$.\myspace
Can we safely compute the real part of the one-loop amplitude within the ET? The technical issue of gauge invariance was already
discussed in the previous subsection. Now we can ask ourselves what the precision of such an approximation is. In the context of partial wave analysis, \textit{perturbative unitarity} relates the
imaginary part of the NLO wave and its leading-order (LO) modulus
\begin{equation}\label{eq: perturbative_unitarity}
Im\,t_{00}^{WW,(4)}=\sigma_W|t_{00}^{WW,(2)}|^2+\sigma_H|t_{00}^{Wh,(2)}|^2.
\end{equation}
These partial waves are defined in detail in the next section.\myspace
Our test consists in computing the left-hand side of (\ref{eq: perturbative_unitarity}) within the ET, via a one-loop calculation,
and checks whether the relation (\ref{eq: perturbative_unitarity}), with the rhs determined using physical $W$'s,
stands and what level of agreement is obtained. For this, let us define the error, in a percentage, assumed by the ET with the quantity
\begin{equation}\label{eq: error_OT}
\Delta_{\text{Full-ET}}=\frac{\bigg |Im\,t_{00}^{\omega\omega,(4)}-\left(\sigma_W|t_{00}^{WW,(2)}|^2+\sigma_H|t_{00}^{Wh,(2)}|^2\right)\bigg |}{\sigma_W|t_{00}^{WW,(2)}|^2+\sigma_H|t_{00}^{Wh,(2)}|^2}\cdot 100.
\end{equation} \myspace
This quantity includes, by construction, couplings from the leading-order Lagrangian (\ref{eq: lag2}), and it is completely independent
of any $\mathcal{O}(p^4)$ parameter since our calculation is made up to $\mathcal{O}(p^4)$ and they just enter at tree level and
consequently do not produce any imaginary part for the left-hand side of equation (\ref{eq: perturbative_unitarity}).
\begin{figure}
\centering
\includegraphics[clip,width=13cm,height=9cm]{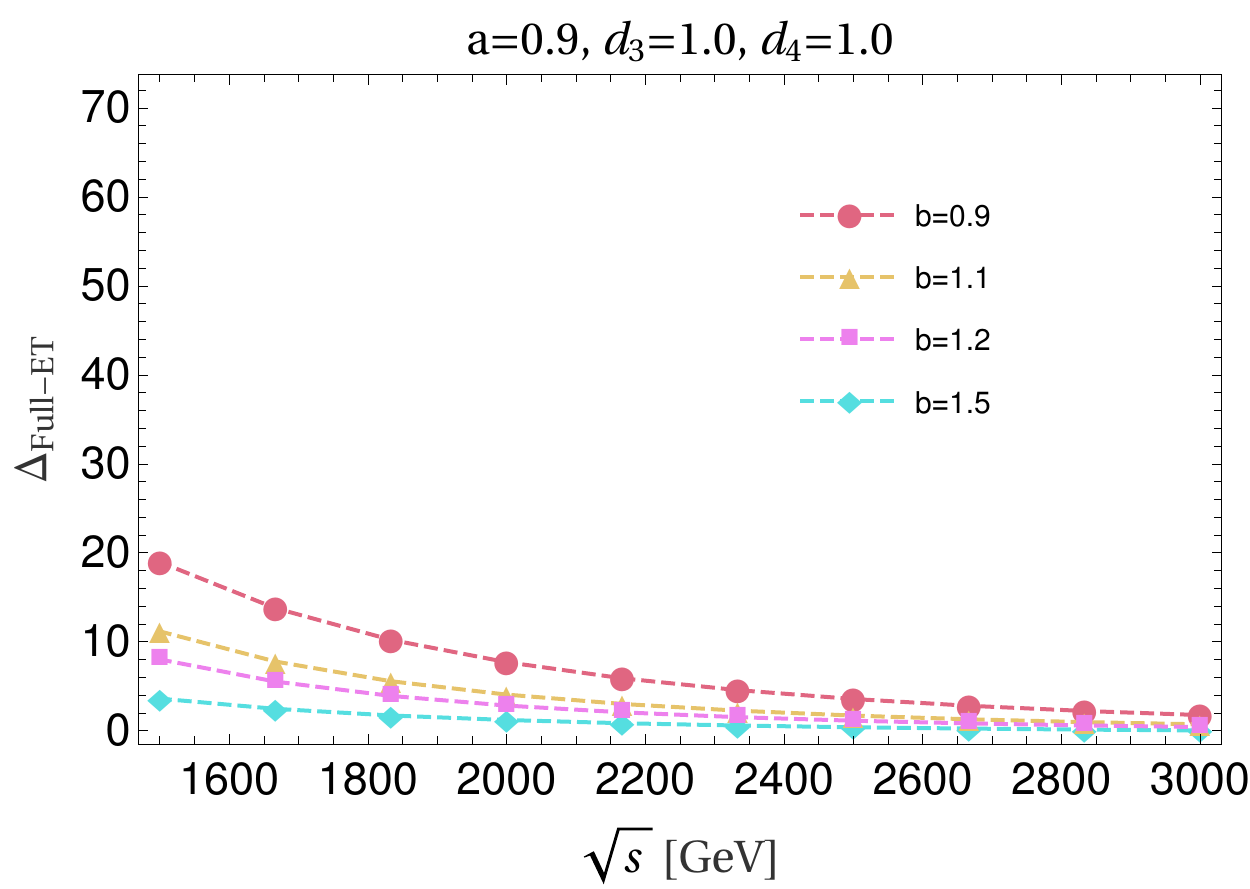}
\caption{\small{In this plot we show in percentage the quantity $\Delta_{\text{Full-ET}}$ defined in (\ref{eq: error_OT}) plotted
    with respect to the centre of mass energy. In this figure we set $a=0.9, d_3=1.0$ and $d_4=1.0$ and show different values of $b$.
    This figure is independent of $\mathcal{O}(p^4)$ parameters. This shows that for large values of $s$ the imaginary part computed
    via the ET agrees at the 10\% level with the one determined (exactly) via the optical theorem. As explained in the text, although the
    discrepancy may look large at low values of $s$, the amplitude is very small there and does not contribute significantly to the position
    and width of possible resonances. See also  Fig. \ref{fig: perturbative_unitarity}.}}
\label{fig: error_OT}
\end{figure}\myspace
Figure \ref{fig: error_OT} shows $\Delta_{\text{Full-ET}}$ for the BSM interaction of a Higgs to two gauge bosons, $a=0.9$, the self-couplings of the Higgs set to their
standard values, and various $b$ values. The behavior is that expected for the ET: the longitudinal components of the electroweak
gauge bosons are well represented by the associated Goldstone boson for high energies compared to the gauge boson masses. This is
independent of the value of $b$ in the plot.\myspace
What we also observe in Fig. \ref{fig: error_OT} is that for values of $b$ close to $a^2=0.9^2=0.81$ the error grows. This "failure"
of the ET can be understood by going to the completely massless limit, useful for $\sqrt{s}\gg M_{\omega,h}$, where the leading-order
$\omega\omega$ and $\omega h$ amplitudes are proportional to $(1-a^2)$ and $(a^2-b)$ [Eqs. (\ref{eq: A_tree}) and (\ref{eq: M_tree})],
respectively. For a fixed value of $a$, close to the SM which cancels the $\omega\omega$ massless amplitude, the closer $b$ is to $0.81$,
the worse the comparison with the full calculation is. This happens because the right-hand side of (\ref{eq: perturbative_unitarity})
approaches zero, being proportional to $b-a^2$, and cancels the leading part of the denominator of $\Delta_{\text{Full-ET}}$. Generally speaking, if one considers the SM parameters $b=a^2=1$, we
do not find good agreement and more terms in the ET expansion, such as $\mathcal{O}(g^2)\,\, W_L\omega\to\omega\omega$, would be
needed at low $s$ values as explained in Ref. \citep{Espriu:1994ep}.\myspace
However, the apparent failure just described is actually a mirage because we are dealing with partial waves that are very small
numerically. To see this,  we show in Fig. \ref{fig: perturbative_unitarity} a check of perturbative unitarity for a benchmark point
away from the SM by a $10\%$, except for the Higgs self-couplings which remain set in their SM values.
\myspace
It could seem that Fig. \ref{fig: perturbative_unitarity} enters into contradiction with the previous Fig. \ref{fig: error_OT} that
shows a worse agreement for the same benchmark point, while it is almost unnoticeable for the low-energy points
in the former. This situation, that is reproduced with any choice of parameters, is explained by the fact that, by construction, the
chiral amplitudes are much smaller at low energies than in the high-energy regime where this uncontrolled growth leads precisely to
the violation of unitarity. Having said that, we can conclude that it is safe for us to make use of the ET for the one-loop level
along the whole range of energies since the differences with the full calculation are on one hand negligible in the high-energy
regime and, on the other hand, the low-energy contributions are much smaller when one is very close to the SM values. Away from the SM limit
(which is of course our main focus), the agreement is good everywhere.

\section{Unitarization}\label{sec:unitarization}
The resonances that we are seeking  cannot be described by a series expansions in the momenta since they should arise as
poles in $S$-matrix elements and that is why we need nonperturbative methods to extend the predictivity of the
low-energy theory, that eventually will lose unitarity, to the strongly interacting regime at higher energies.\myspace
The loss of unitarity requires unitarization methods and there is a variety of such methods (K-matrix, N/D,
IAM, etc: see Ref. \cite{Garcia-Garcia:2019oig} for a complete summary) but they have been proven to show the same qualitative results.
They are based on partial wave analysis and make use of amplitudes with fixed spin $(J)$ and isospin $(I)$ after the projection
\begin{equation}\label{eq: tij_integral}
t_{IJ}(s)= \frac{1}{32K\pi}\int_{-1}^{1}d(\cos\theta)P_J(\cos\theta)T_I(s,\cos\theta)
\end{equation}
where $K$ is a constant whose value is $K=2$ or $1$
depending on whether the particles participating in the process are identical or not and $T_I$ are the fixed isospin
amplitudes that are built from Feynman diagrams and weak isospin relations. These are Eqs.  (\ref{eq: isos_I1})
to (\ref{eq: isos_T0}) of Sec. \ref{sec:amplitudes}.\myspace
As we did in the vector case, we will assume that the scalar wave admits an
expansion in powers of the momenta
\begin{equation}\label{eq: t00_expansion}
t_{00}=t_{00}^{(2)}+t_{00}^{(4)}+\cdots
\end{equation}
restricting ourselves to the lowest order $(I,J=0,0)$ for the study. Hence, with this analysis, we will just be looking for
scalar-isoscalar resonances.

\subsection{Coupled channel formalism}
In our preceding study \cite{Asiain:2021lch} we focused on the case of vector-isovector resonances present in the
$W_LW_L$ elastic scattering, so the
single-channel formalism of the IAM was the way to go. As already anticipated in the Introduction, for the scalar case we
 must take into account the presence of scalar waves coming from double-Higgs configurations (in addition to the $I=0$ projection
of $\omega\omega$) in the intermediate states
of the resummation of bubble diagrams. This mixing of different possible intermediate states is represented in the matrix form of the
scalar-isoscalar partial wave
\begin{equation}\label{eq: t00_matrix}
t_{00}=\begin{pmatrix}t_{00}^{WW} & t_{00}^{Wh} \\ t_{00}^{Wh} & t_{00}^{hh}\end{pmatrix}
\end{equation}\myspace
that is the fundamental structure that will be rendered unitary.\myspace
For the case $b=a^2$ and in the high-energy limit where the mass of the Higgs can be neglected, the off-diagonal elements
(what we call the crossed channel) of (\ref{eq: t00_matrix}) vanish
in the ET limit when we set $g=0$, i.e., in the nET framework. This actually leads to
the decoupling limit: there is no mixing among the different scalar channels whatsoever. However, this is not true as soon as we
set $g\neq 0$, even if $b=a^2$, and the full coupling matrix needs to be considered.\myspace
It can be found, for example in Ref. \cite{Oller:1998hw}, that when cutting the expansion of the scalar wave at NLO [$\mathcal{O}(p^4)$], the
multichannel IAM amplitude is just the generalization of the elastic case in matrix form
\begin{equation}\label{eq: t00_IAM}
t_{00}^{IAM}=t_{00}^{(2)}\cdot\left(t_{00}^{(2)}-t_{00}^{(4)}\right)^{-1}\cdot t_{00}^{(2)}
\end{equation}
The elements of the IAM matrix are all the unitary scalar waves participating in the process up to NLO: unitary $WW$ in the first diagonal
entry, $Wh$ in the off-diagonal and $hh$ in the second diagonal element.\myspace
This IAM matrix, besides keeping the analytical properties on the right cut required for partial wave analysis, has a low-energy
expansion that coincides with (\ref{eq: t00_expansion}) and fulfills the exact unitarity condition
\begin{equation}\label{eq: exact_unitarity}
Im\,t_{00}^{IAM}=\sigma\left(t_{00}^{IAM}\right)^{\dagger}t_{00}^{IAM}.
\end{equation}
where $\sigma$ is the two-body phase space. At this point, we find an ambiguity in the crossed channel of this expression: we have
two kinds of particles with different masses, the gauge bosons and the Higgs, but yet we only include a unique phase space, that we
choose to be the one with the $W$ boson mass, $\sigma=\sqrt{1-\frac{4M_W^2}{s}}$. This choice, of course, will be of no relevance at
the high-energy regime that we want to explore where $M_W^2\approx M_H^2\ll s$ and $\sigma \approx 1$.\myspace
From (\ref{eq: t00_IAM}), it can be seen how the scalar resonances, if present, are located at the zeros of the determinant
\begin{equation}\label{eq: t00_determinant}
\Delta_{00}(s_R)\equiv\det\left(t_{00}^{(2)}(s_R)-t_{00}^{(4)}(s_R)\right)=0
\end{equation}
where the Breit-Wigner resonances occur at $s_R=\left(M_R\ih\Gamma_R\right)^2$ in the $s$-complex plane.\myspace
With this coupled-channel formalism, more channels are available for the resummation of the intermediate and in the final
states, making the resonances appearing in the scattering characteristically broader. They are short lived, compared to
those found in single-channel, massive states.
If these poles in the zeros of the determinant (\ref{eq: t00_determinant}) are to be interpreted as Breit-Wigner-like resonant states,
we will be applying the broadly used criterion that the width satisfies $\Gamma<\frac{M}{4}$, meaning this that the pole
is located near the real axis as one can see from the definition of $s_R$ above. Otherwise, we would have found a simple
enhancement of the scalar amplitude not to be interpreted as a physical pole with such an enhancement produced by the presence
of a pole far from the real axis.\myspace
As in the vector case, we shall be looking for poles appearing in the second Riemann sheet where the Breit-Wigner interpretation
leads to positive widths, required by causality arguments. If some pole appears in the first Riemann sheet, where imaginary parts
are positive, it would be associated with a spurious resonance with negative width that cannot be present in a physical theory.
Thus, we find here an empirical approach in order to discriminate {\em a priori} plausible sets of parameters in the HEFT.\myspace
\subsection{Scalar resonances}
\begin{figure}
\centering
\includegraphics[clip,width=8.1cm,height=6.3cm]{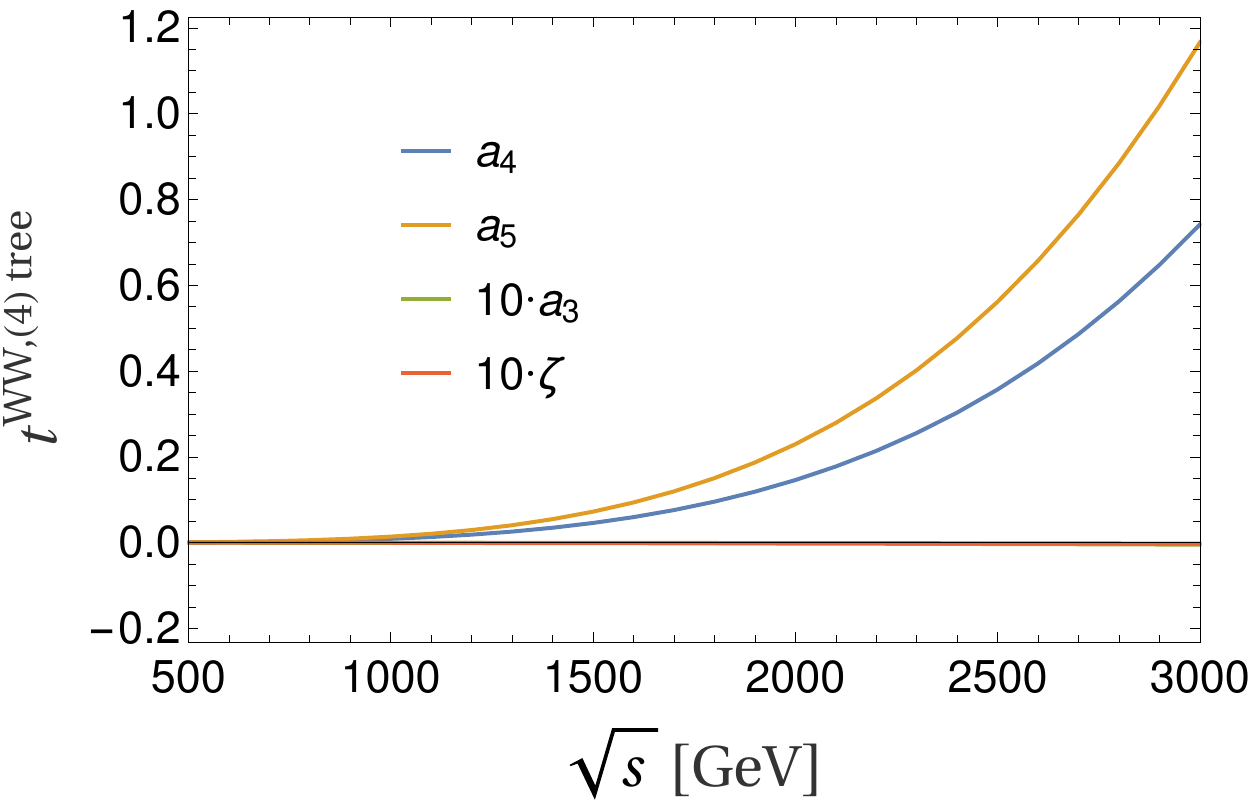} %
\includegraphics[clip,width=8.1cm,height=6.2cm]{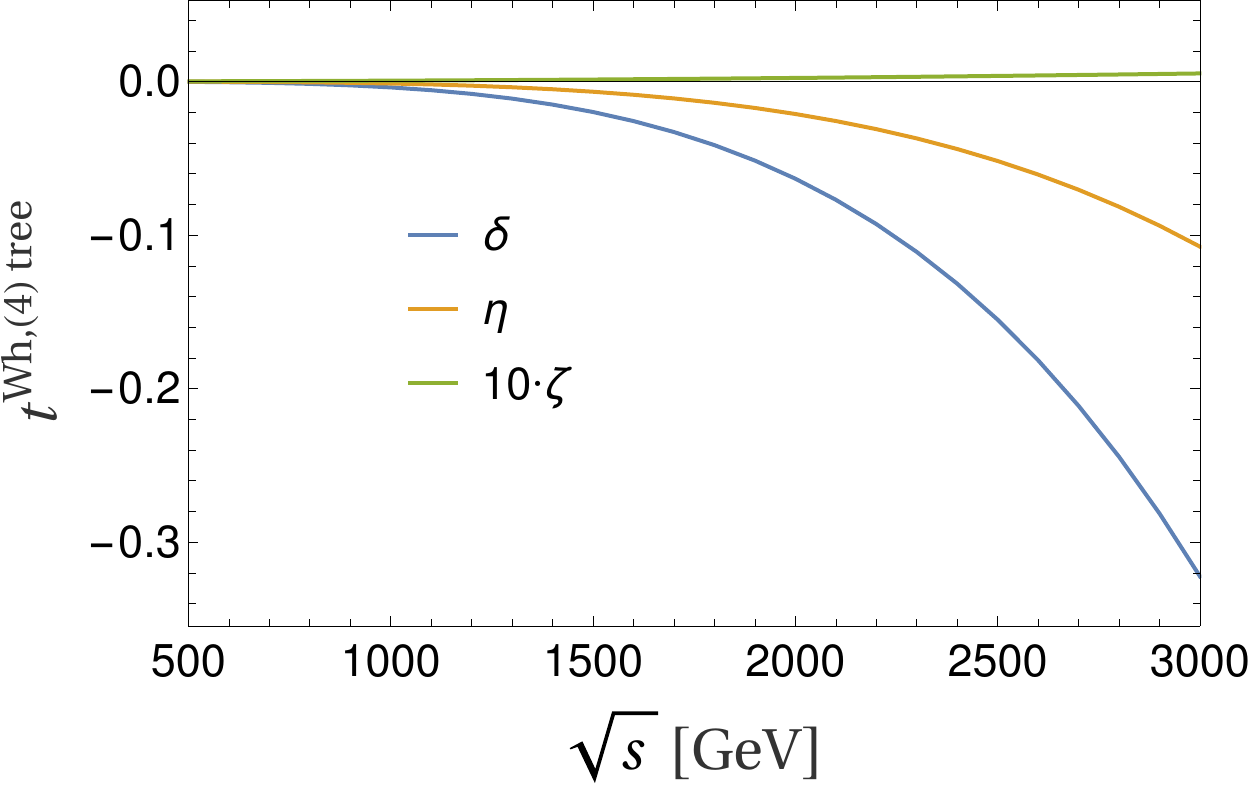} %
\caption{\small{Plot of the NLO tree-level scalar wave separated in the different chiral coupling contributions for (left axis) 
the elastic $WW$ and (right axis) the crossed channel $Wh$. All the values are chosen to be of the maximum expected size 
of $10^{-3}$.}}%
\label{fig: relative}
\end{figure}

\begin{figure}
\centering
\includegraphics[clip,width=13cm,height=9cm]{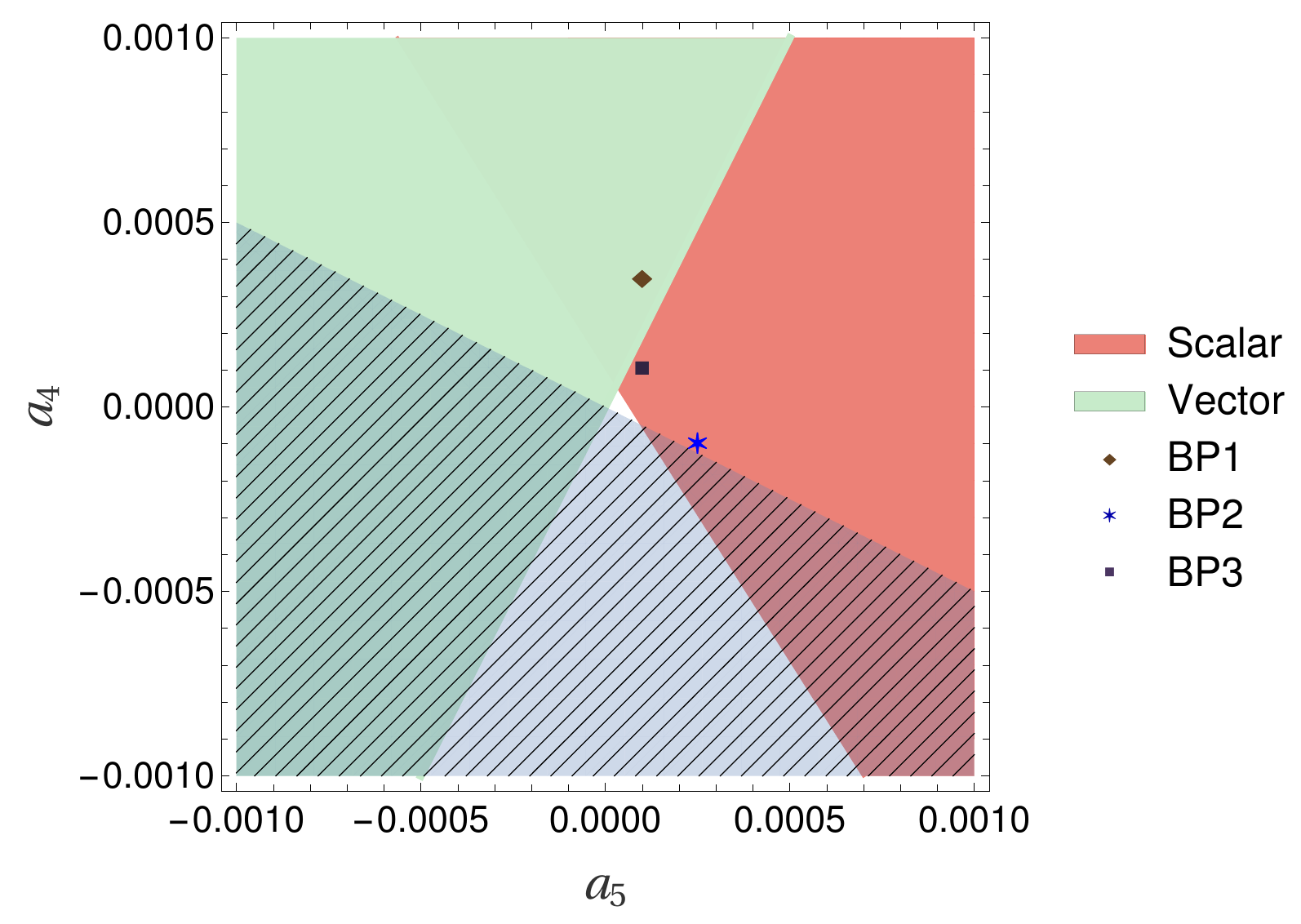} %
\caption{\small{Regions in the $a_4-a_5$ plane where scalar (red) and vector (green) resonant states appear. The striped area represents excluded paramter space by the presence of acausal isotensor resonances. The benchmark points used in this
    study are marked  in the plot. One of them, BP1, lies in the region where both isoscalar and isovector states show up and the other two, BP2 and BP3,
    in the purely isoscalar sector.}}
\label{fig: map}
  \end{figure}

Now that we have presented the coupled-channel formalism for the scalar waves, we are prepared to search for scalar resonances
in the chiral parameter space.\myspace
Following the way of building the fixed $I=0$ isospin amplitude in Ref. \citep{Asiain:2021lch}, the scalar partial wave is obtained
from (\ref{eq: tij_integral})
\begin{equation}\label{eq: t00_integral}
t_{00}=\frac{1}{64\pi}\int_{-1}^{1}d(\cos\theta)T_{0}(s,\cos\theta)
\end{equation}
where we have used that the Legendre polynomial $P_{0}(\cos\theta)=1$.\myspace
Previous works such as Refs. \citep{Delgado:2014dxa} and \citep{Arnan:2015csa} already searched for scalar resonances in $WW$ scattering
following the procedure developed in the preceding section. These works relied on the ET (even at tree level) in the naive custodial
limit, $g=g^{\prime}=0$, and the former assumed a completely massless scenario, so both of them could get exact analytical
continuations of the partial waves to the second Riemann sheet to look for the resonant states. This is a step that we are not
able to perform in our calculation, as the resulting expressions do not have an analytic treatment.\myspace
The first task for our numerical analysis, following the ideas in Ref. \citep{Asiain:2021lch}, is to find modifications
in the properties of the scalar resonances studied e.g., in Ref. \citep{Arnan:2015csa} once one relaxes the $g=0$ approximation and includes
gauge bosons in the external states at tree level and in internal lines of the one-loop calculation. In that study, the authors
considered the relevant chiral parameter space giving scalar resonance masses in the range $1.8$ TeV$<M_S<2.2$ TeV. No coupled-channel
formalism was used, and instead they assumed the decoupled-channel limit within the nET by setting $b=a^2$ for the particular
case $b=a=1$. For some benchmark points in the mentioned region, we get the modifications on the location of the scalar points after allowing for transverse gauge propagation (see Table \ref{table: differences_arnan})
\begin{table}[h!]
\begin{tabular}{|c|c|c|c|c|}
\hline
    $\sqrt{s_S} \, (GeV)$ &$a_4\cdot 10^4$ & $a_5\cdot 10^4$ & $g=0$ & $g\neq 0$\\ \hline
$ $  & $ \quad 1 \quad $ &     $\quad -0.2 \quad $    & $\quad 1805 \ih 130 \quad $ & $\quad 1856 \ih 125\quad$            \\ \hline
$ $  & $\quad 2 \quad$ &   $\quad -1  \quad$      & $\quad 2065 \ih 160 \quad$ & $\quad 2119 \ih 150 \quad$\\ \hline
$ $ & $\quad 3.5 \quad $ &  $\quad -2 \quad$       & $\quad 2175\ih 170 \quad$ & $\quad 2231 \ih 163 \quad$ \\ \hline
\end{tabular}
\caption{{\small
    Values for the location of the scalar poles $\sqrt{s_S}=M_S-\frac{i}{2}\Gamma_S$ for $g=0$ and $g\neq 0$ for
    some points in the $a_4-a_5$ plane
    and in the decoupling limit $b=a^2$ within the nET with $a=b=1$. The self-interactions of the Higgs are set to the
    SM values. Note that the coupling to other $I=0$ channels is ignored here for the purpose of
    assessing the effect of switching on the transverse modes.}}\label{table: differences_arnan}
\end{table}\myspace
From this table we extract similar conclusions as in the vector case. On one hand, the masses of the scalar resonances are
pushed up by $2\%-3\%$ once the $SU(2)_L$ coupling is set to its SM value, the very same behavior as for $I=1$. On the
other hand, we observe variations in the widths of around $4\%-6\%$, values much greater than in the vector case where the differences
were almost unnoticeable. This gives us an idea of the significance of the propagation of transverse modes.\myspace
However, the above results for $M_S, \Gamma_S$ are only tentative because when one wants to make a full calculation beyond
the nET and consider physical vector bosons in the external states, even in the case where $b=a^2$, there is no decoupling
and one needs the coupled-channel formalism to get a proper description of the dynamics of the system in the $IJ=00$ channel.
Let us now proceed to study how coupling the various relevant channels affects the results. The modifications will be
substantial in fact.\myspace
All the $\mathcal{O}(p^2)$ parameters are included in all the amplitudes of (\ref{eq: t00_matrix}), but as one can see from the 
$\mathcal{L}_4$ Lagrangian (\ref{eq: lag4}), not every $\mathcal{O}(p^4)$ coupling affects all the channels. In particular, 
$WW$ depends on $a_4$, $a_5$, $a_3$, and $\zeta$; $Wh$ depends on $\delta$, $\eta$, and $\zeta$; and the $hh$ elastic process depends only on $\gamma$.
The operators accompanying these couplings could eventually dominate the corresponding amplitudes at high energies due to 
the presence of the four derivatives. However, not all these couplings contribute to the NLO scalar amplitude with the same strength. 
The aforementioned contribution is represented in Fig. \ref{fig: relative} for values of the parameters of the expected (absolute) 
size of $10^{-3}$. 
\myspace
From Fig.\ref{fig: relative}, we see an evident hierarchy among the different couplings: $a_4$ and $a_5$ contributions are much 
more relevant than those of $a_3$ and $\zeta$ in the elastic $WW$. For the crossed scattering, $Wh$, $\delta$, and $\eta$ 
contributions are much more important than the one of $\zeta$. This picture reinforces the conclusion that those operators 
surviving in the $g=0$ limit (the nET limit) are more relevant that the other ones. The reason why this happens lies in the fundamental
structure of the HEFT. To be consistent in the chiral counting, both Higgs mass ($\sim \sqrt{\lambda_{SM}} v$) and EW gauge boson mass ($\sim gv$)
must be understood as $\mathcal{O}(p)$ soft scales; therefore, a local operator with one gauge coupling plus three derivatives
(like those accompanying $a_3$ and $\zeta$) is of chiral order $4$, just like one with four derivatives ($a_4, a_5,\cdots$), but the
latter dominates by far at high energies. \myspace
The behavior presented above agrees with what we found from vector-isovector resonances \citep{Asiain:2021lch}: the pole position was
almost completely determined by $a_4$ and $a_5$ with subleading 
effects after adding $a_3$ and $\zeta$, at least for values of $a,b, d_3$, and $d_4$ close to the SM values. This is why in the
forthcoming analysis, in order to keep it as simple as possible, we will only
consider the influence of $a_4$ and $a_5$ in determining the properties of resonances in the $IJ=11$ channel and neglect
the role of $a_3$ and $\zeta$.\myspace 
The space of parameters to analyze in the $IJ=00$ case is considerably larger than in the vector case and some sort
of hierarchy is needed in order to proceed. One point to check is whether in the scalar case $a_4$ and $a_5$ dictates
to a very good approximation the structure of resonances as it happens in the vector case (assuming for the time being that
we stay close to the SM values $a=b=d_3=d_4=1$). To study this,  we will focus 
first on the benchmark points (BPs) in the $a_4-a_5$ plane defined in Table \ref{table: BPs}. Other works have studied the spectrum of
resonances in $WW$ scattering, in particular the group of the reference \citep{Rosell:2020iub} that made use of Weinberg
sum rules Ref. \citep{Weinberg:1967kj} derived from the $W^3B$ correlator, to set minimal bounds for the masses of vector resonances
allowed by experimental constraints of the chiral parameters. For the region in the $a_4-a_5$ plane that we are interested in,
they found that, for any scenario where an axial state is decoupled, the minimal mass for an experimentally allowed vector
resonance is around $2$ TeV. We slightly relax that condition and require a parameter space where, if present, the vector resonances satisfy $M_V\gtrsim 1.8$ TeV. We choose the minimal mass for any observable scalar resonance to be the same value of $M_S\gtrsim 1.8$ TeV and assume that any lighter state should have already been seen in the experiment.\myspace
At this point, one should recall that only particular combinations of $a_4$ and $a_5$ appear in the various channels, namely, $5a_4+8a_5$ for $IJ=00$, $a_4-2a_5$ for $IJ=11$, and $2a_4+a_5$ for $IJ=20$ \cite{PhysRevD.55.4193}. In previous studies, it was found that
isotensor resonances are always acausal and the corresponding region $2a_4+a_5<0$ is to be excluded from our considerations. Thus, we select BPs outside the region excluded by isotensor acausal resonances \cite{Espriu:2014jya} and within the vector-isovector and
scalar-isoscalar space. In particular, we select one BP (BP1) that belongs to the region where both isovector and
isoscalar resonances appear, satisfying the condition commented above regarding the vector resonance mass. The other two BPs (BP2 and BP3) lie
in the purely scalar-isoscalar region. In Fig. \ref{fig: map}, we see the position of the BPs that we have just described within the $a_4-a_5$
plane where resonances are present when the coupled-channel formalism is applied. \myspace
\myspace
These BPs are gathered in Table \ref{table: BPs} where we also include, even if they do not have a physical relevance,
the values after applying the single channel formalism to the $WW$ scalar
wave, obviating the crossed channel and the elastic $hh$ scattering. Both values are obtained with $g\neq 0$.\myspace
\begin{table}[h!]
\begin{tabular}{|c|c|c|c|c|c|}
\hline
$  $ & $ a_4\cdot 10^4 $ & $a_5\cdot 10^4$ &  $S.C. $ & $C.C. $ & $M_V\ih \Gamma_V$\\ \hline
$ BP1$ & $ 3.5  $ & $ 1 $   &  $ 1044\ih 50  $ & $ {\bf 1844\ih 487} $ & $2540\ih 27 $ \\ \hline
$ BP2$ & $ -1 $  & $ 2.5 $   &  $ 1219\ih 75 $ & ${\bf 2156\ih 637} $ & $\redcross$ \\ \hline
$ BP3$ & $ 1 $  & $1 $   &  $1269\ih 75 $ & ${\bf 2244\ih 675} $ & $\redcross$ \\ \hline
\end{tabular}
\caption{{\small Properties of the scalar resonances for the selected benchmark points in the $a_4-a_5$ plane, with
    the $\mathcal{O}(p^2)$ parameters set to their standard values, in both single-channel (S.C.) and coupled-channel (C.C.)
    formalism. We also include the values of the properties of
    vector resonances if present. The centerdots in red $\redcross$ indicates the absence of a zero in the
    determinant, Eq. (\ref{eq: t00_determinant}).
    The $\mathcal{O}(p^2)$ chiral parameters are set
    to their SM values. We see that coupling channels modifies very substantially masses and widths. Those poles not
    fulfilling the resonance condition are in boldface.}}\label{table: BPs}
\end{table}\myspace
The first thing that one notices is that when (correctly) considering coupled channels the results differ considerably from
the ones one would obtain in the single channel and the resonance masses and widths visibly increase. Recall that here we are
assuming $a=b=1$ where naively one would expect
to have decoupling (this is the case in the nET). This is not so because we are setting $g\neq 0$. In fact some of the would-be
resonances even dissapear as such by just becoming broad enhancements. Recall that conventionally a physical resonance must
satisfy $\Gamma<M/4$ and this is not the case in many cases when applying the coupled-channel formalism. Obviously,
coupled channels matter.\myspace
Finally, let us mention that in $pp$ collisions the scattering of vector bosons (VBS) is a subdominant process, but the production
of $hh$ pairs via VBS is further suppressed with respect to the elastic channel $WW \to WW$. A relevant issue that will be studied
below is the intensity of the coupling of the dynamical resonances to $hh$ final states. As we will see, they would be more visible
in the elastic $WW\to WW$ channel and tend to couple weakly to final $hh$ pairs. To what extent this depends on the various couplings is an 
interesting question, too.\myspace
\subsection{Checking unitarity}
\begin{figure}
\centering
\includegraphics[clip,width=8.1cm,height=6.8cm]{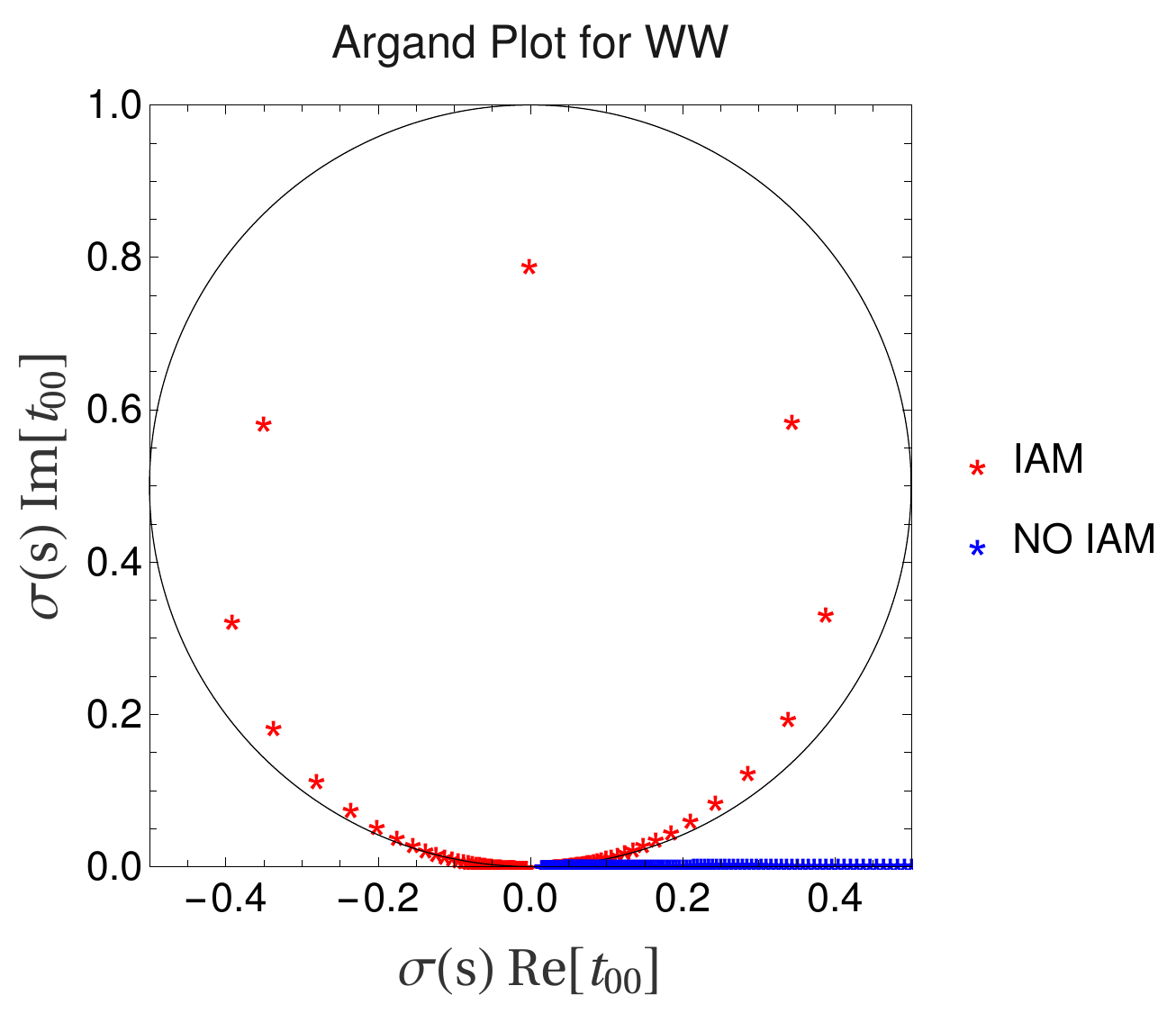} %
\includegraphics[clip,width=8.1cm,height=6.8cm]{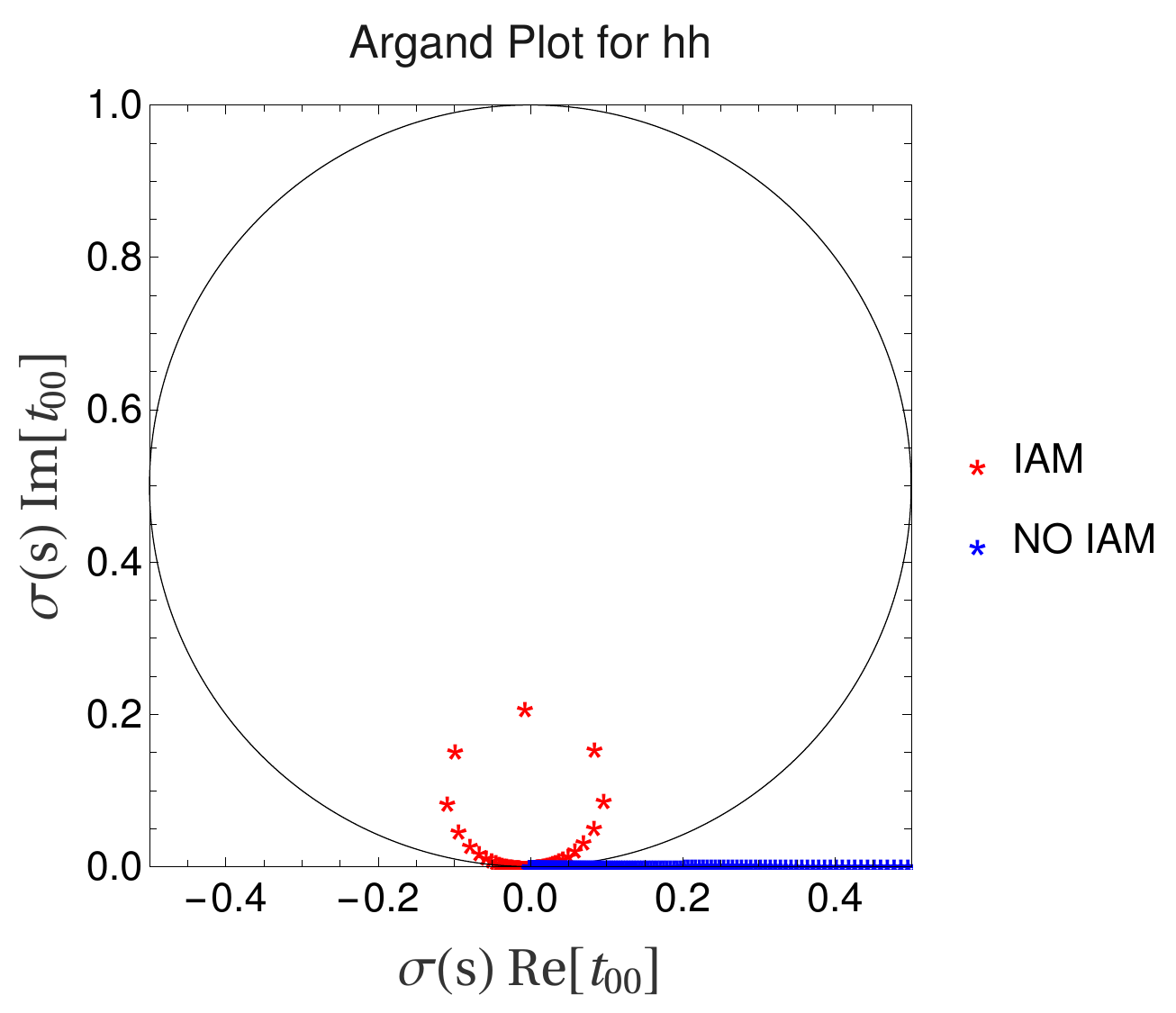} %
\caption{\small{Argand plot for the scalar wave of the (Left) elastic $WW$ and (Right) elastic $hh$ scattering for BP3 and the chiral
    coupling $\gamma=10^{-3}$. The rest of the parameters are set to their SM values. In red dots the unitarized amplitude satisfying the
    unitarity condition and in blue dots the non-unitary chiral amplitude from the Lagrangians (\ref{eq: lag2}) and (\ref{eq: lag4}).
    The $Wh$ crossed channel alone needs not to satisfy this condition of lying on or inside the circumference.}}
\label{fig: argand}
  \end{figure}

The bad high-energy behavior, manifest in the amplitudes of the effective theory even,
must be avoided in order to give reasonable predictions that do not overestimate the number of predicted events
in $WW$ fusion subprocesses. Here, we provide evidence that this is the case when the partial scalar waves are unitarized.\myspace
Indeed, the IAM amplitude (\ref{eq: t00_IAM}) is built to keep the desired unitarity property (\ref{eq: exact_unitarity}) as
long as there is a good
description of the amplitudes across the physical cut. One can easily show from the exact unitarity condition that the unitarized partial
waves for the elastic processes ($WW$ and $hh$) must lie on or within the circumference of radius $1/2$ centered at $(0,1/2)$ when plotting
the imaginary versus the real part of the unitarized amplitude. This is not the
case for the crossed-channel $Wh$. As a demonstration of the good behavior of our scalar waves, we show in Fig. \ref{fig: argand} the
Argand plot for the elastic processes described with the chiral parameters of BP3 plus $\gamma=10^{-3}$. The rest of the chiral couplings
keep their SM values.\myspace
As we can see in Fig. \ref{fig: argand}, no matter the point (every one of them corresponding to different energies), they all fall within
the unitary circle. The fact that they do not lie exactly over the circumference is because, for the selection of parameters chosen for
illustration, there is a big component of inelasticity in the process, i.e., the crossed channel cannot be neglected.
\myspace
\subsection{Some theoretical insight}
Physical resonances should always be located in the second Riemann sheet. 
It is sometimes not fully appreciated that the presence of unphysical singularities, e.g., in the first Riemann sheet,
is useful to restrict the
space of parameters of an effective field theory. In hadron physics, for instance, it is widely known that a broad range of parameters
in the pion Effective Chiral Lagrangian is excluded because it leads to acausal resonances in isotensor resonances.
This result automatically translates into the HEFT because the expressions are very similar. An upper bound on the
combination $a_5+ 2a_4$  (see Ref. \citep{PhysRevD.55.4193}) emerges.
This restricts the range of parameters that can be considered for an effective theory. In other word, no UV completion may
exist that leads to such low-energy constant.\myspace
This lack of causality can also be understood directly on Lagrangian terms in some cases.
In Ref. \cite{Adams:2006sv}, it was seen in a general setting how such restrictions may arise.\myspace
In the context of HEFT, in Ref. \cite{Urbano:2013aoa}, the following sum rule was derived
\begin{equation}\label{eq: sum_rule}
\frac{1-a^2}{v^2} = \frac{1}{6\pi}\int_0^\infty \frac{ds}{s}
\left(2\sigma_{I=0}(s)^{tot} + 3 \sigma_{I=1}(s)^{tot} -5
\sigma_{I=2}(s)^{tot}\right),
\end{equation}
where $\sigma_I^{tot}$ is the total cross section in the isospin channel $I$. This interesting result
was derived making full use of the ET and setting $g=0$. Taking into account that unless there is an unlikely
strong enhancement of the $I=2$ isospin channel, the rhs is positive definite, this would exclude
values of the effective coupling $a$ greater than 1. Note that we just saw that there were no physical resonances in the isotensor
channel. And, indeed, no satisfactory microscopic model has
been constructed with $a>1$ to our knowledge.\myspace
As we have seen, there are some deviations with respect to the ET predictions when using
the proper longitudinal vector boson amplitudes and they affect the analytic properties of the amplitude.
In Ref. \citep{Espriu:2014jya} it was seen that a complete calculation (as opposed to the simpler nET treatment) changes the previous result
in several ways. For instance, it is not true that a
given order in the chiral expansion corresponds
to a definite power of $s$ ---a property that is used in order to derive (\ref{eq: sum_rule}). Therefore,  when gauge transverse propagation
$q$ is included, the order $s$ contribution will have corrections
from all orders in perturbation theory. The contribution to the left-hand side of the integral obtained will then be of the form
\begin{equation}\label{eq: cont_sum_rule}
\frac{3 - a^2 + \mathcal{O}(g^2)}{v^2}\,.\end{equation}
However, the right cut changes, too, when $g$ is taken to be nonzero due to $W$ propagation in the $t$ channel and this 
could compensate the modification on the lhs.
Finally, as we have seen, crossing symmetry is not manifest in the Mandelstam variables when one moves away from the nET. This is again
a necessary ingredient to derive (\ref{eq: sum_rule}).\myspace
These subtleties, however, do not mean that the $a>1$ forbidden region is not present; it just means that proving this when the propagation
of transverse modes is taken into account is not so easy. Indeed, Ref. \citep{Espriu:2014jya}, it was seen that for $a>1$ the IAM led to pathologies
in resonances appearing
in various channels, including acausal resonances ---poles in the first Riemann sheet.\myspace
Therefore, it seems that an efficient way of setting bounds on the low-energy constants is provided by discarding those regions
of parameter space in the effective theory, i.e., in the infrared, where resonances are acausal. The regions described by these
effective theories do not have an ultraviolet completion.\myspace

\section{Results}
\begin{figure}
\centering
\includegraphics[clip,width=7.0cm,height=6.5cm]{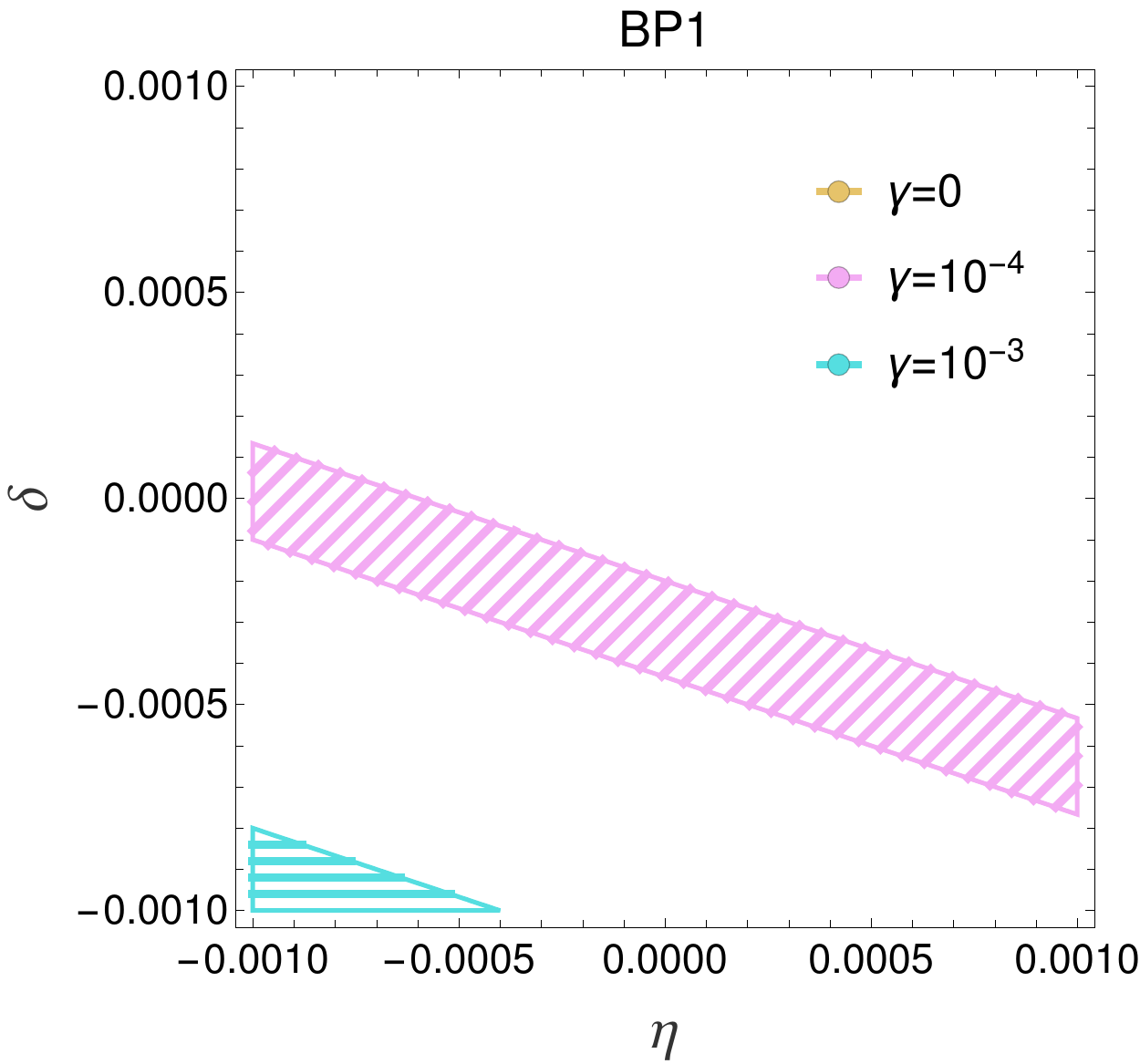} %
\includegraphics[clip,width=7.0cm,height=6.5cm]{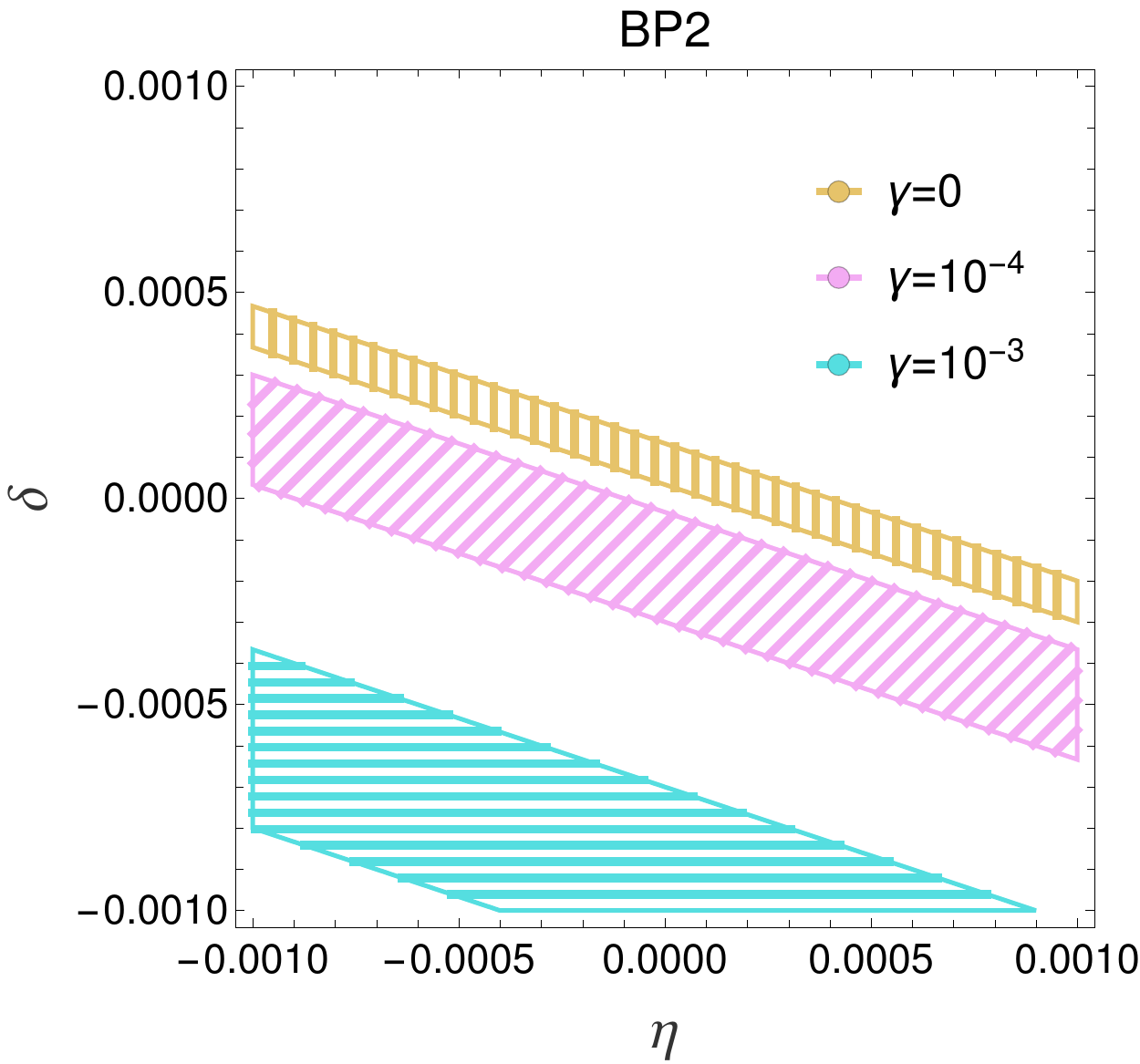} %
\includegraphics[clip,width=7.0cm,height=6.5cm]{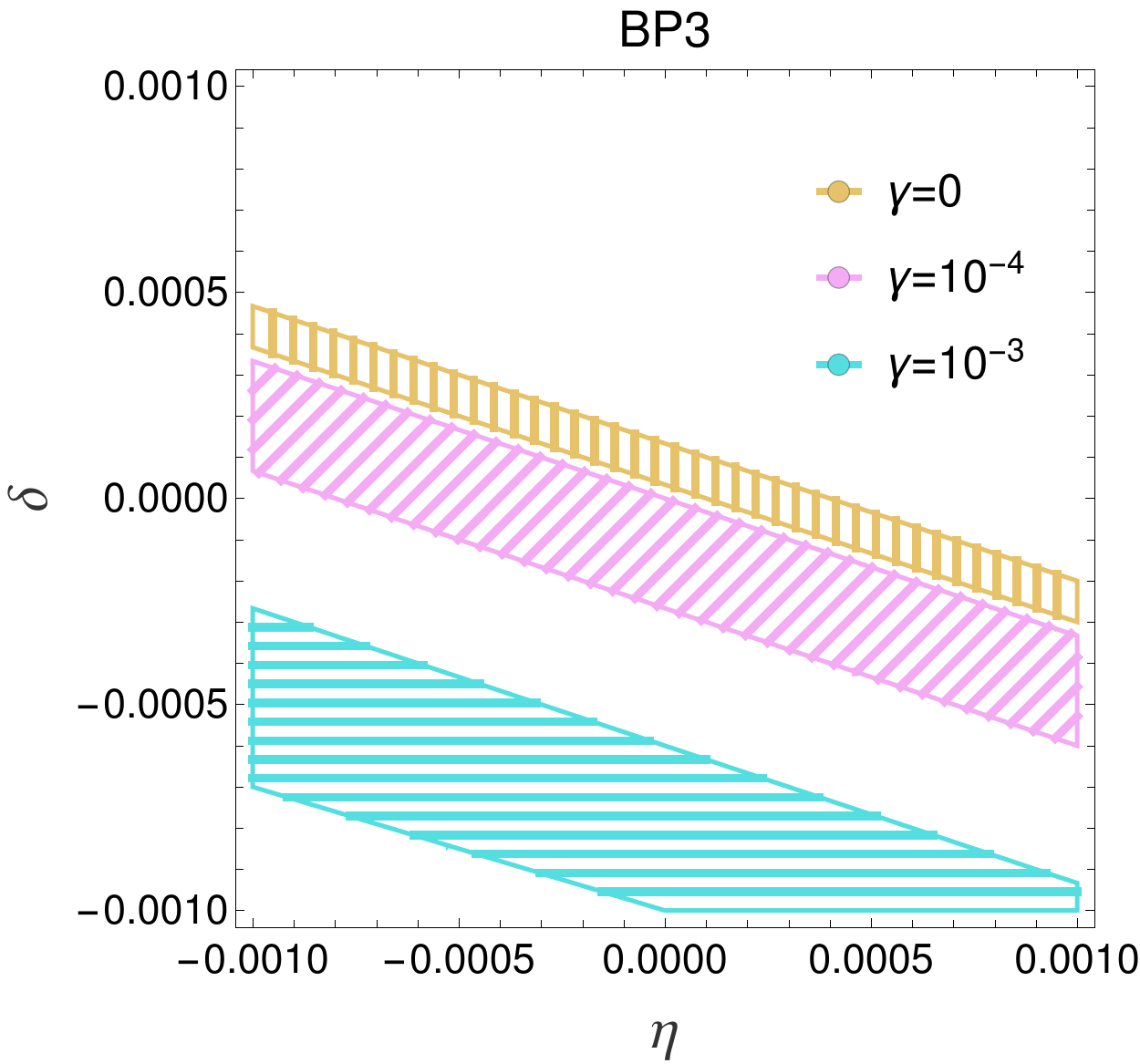} %
\caption{\small{Regions in the $\delta-\eta$ plane where physical resonances satisfying $M_S>1.8$ TeV and for the benchmark points in
    Table \ref{table: BPs} appear for
    different values of $\gamma$: $\gamma=0$ (golden vertical lines), $\gamma=10^{-4}$ (pink tilted lines) and $\gamma=10^{-3}$ (blue horizontal lines). For all the values of $\gamma$, the region above the bands are excluded by the presence of a non-physical pole. Below the bands we find a non-resonant scenario.}}%
\label{fig: swept_gamma}
\end{figure}

\begin{figure}
\centering
\includegraphics[clip,width=7.8cm,height=6.5cm]{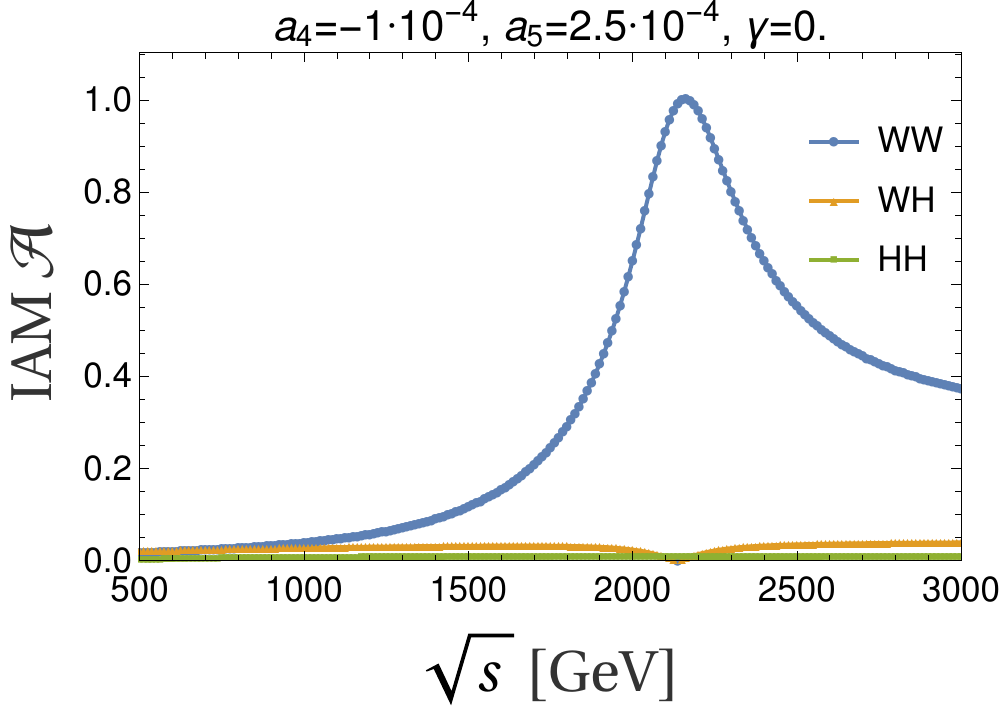} %
\includegraphics[clip,width=7.8cm,height=6.5cm]{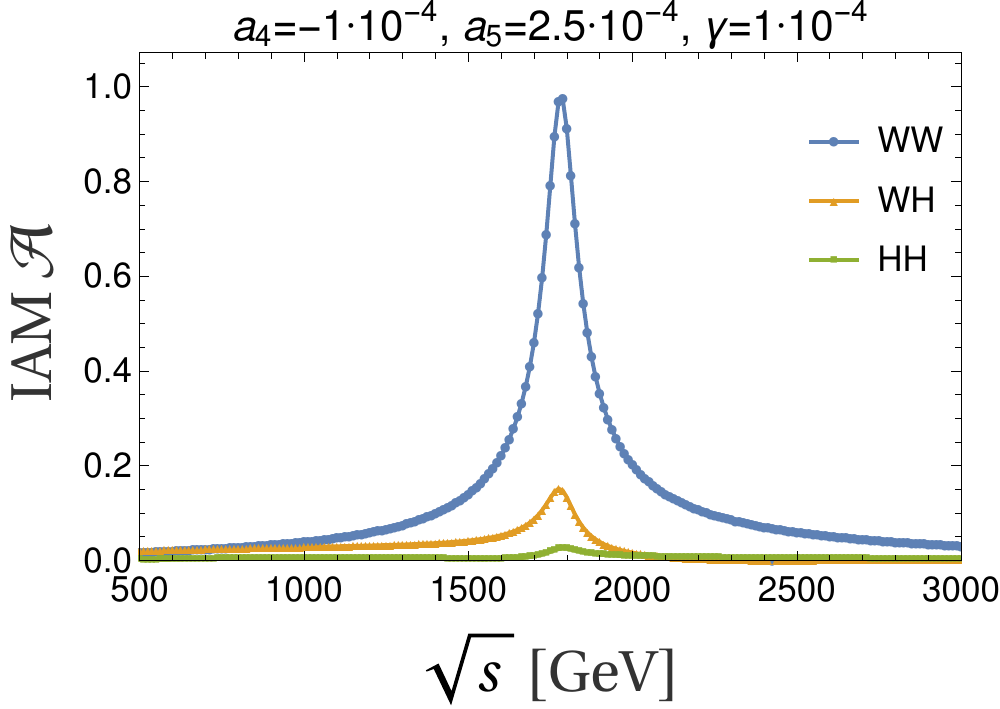} %
\caption{\small{Profile of the unitarized amplitude showing a zero in the determinant for the chiral couplings specified
    in the title and with the rest of the parameters set to the corresponding SM values.}}%
\label{fig: relative_residues}
\end{figure}

As previously  mentioned, bounds on the parameters of the HEFT from the study of unitarity and resonances can come in two ways. One is
simply by experimentally falsifying a given set of parameters because they should give rise to resonances that are not seen in
experiment. The other is giving rise to unphysical acausal resonances that lay on the wrong Riemann sheet.\myspace
To exploit all the potential of the analysis, we will group the parameters of the HEFT into two different sets. One of them contains all
parameters that enter the ${\cal O}(p^2)$ Lagrangian; namely $a,b,d_3$, and $d_4$. In the preliminary analysis previously presented, these values
were all set to their SM values $a=b=d_3=d_4=1$.  The other set contains all the ${\cal O}(p^4)$ parameters: $a_4, a_5, \gamma, \eta$, and $\delta$.
We shall assume that none of the parameters in the second group exceeds in absolute value $10^{-3}$. We will not include the chiral
parameters $a_3$ and $\zeta$ because in Ref. \citep{Asiain:2021lch} it was demonstrated that they play only a marginal role in the determination of
vector
resonances.\myspace
In what follows, we will first study the influence of the relevant ${\cal O}(p^4)$ parameters while keeping the first set to their
SM values. Later, we will repeat the analysis for values of $a,b$ that slightly differ from the SM, but still keeping $d_3=d_4=1$. Finally, we will
study the influence of $d_3$ and $d_4$, but keeping the SM values $a=b=1$ to test the sensitivity of scalar resonances to the
parameters in the Higgs potential.\myspace
For a given set of $\mathcal{O}(p^2)$ parameters ($a,b,d_3$, and $d_4$), once one fixes $a_4$ and $a_5$, the pole position 
in the elastic $WW$ channel is pretty much determined (up to small $a_3$ and $\zeta$ corrections that we neglect). To the extend
that the elastic channel may dominate resonance production, we can treat the effect of the rest of parameters that participate
in the mixing among the scalar channels
as a perturbation. We have, for instance, searched for resonances in the case $a_4=a_5=0$ while varying the remaining
${\cal O}(p^4)$ terms with a negative result. The presence of resonances
(both in the vector and scalar channels) is largely triggered by nonzero values of the chiral couplings $a_4$ and $a_5$.\myspace
However, not every set of low-energy parameters may correspond  to an effective  description of a strongly interacting theory.
Therefore, we have to be able to discriminate which of the zeroes of Eq. (\ref{eq: t00_determinant}) is a physical and which is not and also
which resonances should have also been observed.\myspace
On one hand, we will be looking for resonant states that satisfy the condition $\Gamma < M/4$. If this is not fulfilled,
we will be talking of an enhancement of the unitarized amplitude but never to be interpreted as a resonant state. In that case, even
if Eq. (\ref{eq: t00_determinant}) has a zero, the parameters $M$ and $\Gamma$ are not directly related the properties
of a Breit-Wigner resonance.
On the other hand, there are zeros that even satisfying the aforementioned condition cannot be taken as physical states since they have
negative Breit-Wigner widths. These spurious states cannot be present in any physical theory. Analytically speaking,
these zeros are found in the first Riemann sheet, above the physical cut in the complex $s$ plane.\myspace
Unlike in the case of the nET, due to the complicated structure of the one-loop amplitude, we cannot perform analytical
continuation to the second Riemann-sheet and find poles analytically. To contour that difficulty we have three
tools at our disposal: (1) comparison with the nET in order to see if a pole represents a modification of a pole previously known
to exist in the simplified model; (2) fitting the partial wave to a two Breit-Wigner resonances, leaving the sign of the width as
a free variable; and (3) checking the behavior of the phase shift across the resonance. Of all three possibilities, tool 1 is not very
informative because, as seen, the modifications with respect the nET are large when coupled channels play a role, tool 2 is quite useful, but tool 3 is the method of choice (particularly when combined with tool 2).\myspace
The phase of an amplitude that contains a physical resonance presents a shift from $\pi/2$ to $-\pi/2$ in the pole position.
The derivative of the phase should always be positive as the expression
$\Gamma\sim \left(\frac{\partial \delta(s)}{\delta \sqrt{s}}\right)^{-1}$,
where $\delta (s)$ represents the phase, can be derived analytically. In this way, we study the causal character
of all resonances found. All cases
are met: one resonance, two physical resonances, and also two resonances where only one of them happens to be physical.\myspace
Let us now proceed with the study in the case $a=b=d_3=d_4=1$.\myspace
To study the impact of the new parameters, we focus on the three BP points defined by specific values of $a_4$ and $a_5$
previously used. The new parameters are:
$\gamma$, that enters in elastic $hh$, and $\delta$ and $\eta$ that carry out the mixing between the two elastic processes as one can see in (\ref{eq: t00_matrix}).\myspace
The effect of each $\mathcal{O}(p^4)$ anomalous coupling is reflected in the Tables \ref{table: differences_coupled_gamma}-\ref{table: differences_coupled_eta} below where we study the
separate influence of each one for the above
benchmark points. In the following analysis we keep the SM values for $a,b,d_3$, and $d_4$.
\begin{table}[h!]
\begin{tabular}{|c|c|c|c|c|c|c|}
\hline
    $M_S\ih \Gamma_S$ & $\gamma=0$  & $\gamma=0.5\cdot 10^{-4}$ & $\gamma=1\cdot 10^{-4} $ & $\gamma=-0.5\cdot 10^{-4}$ & $\gamma=-1\cdot 10^{-4}$ & $\gamma=1\cdot 10^{-2} $ \\ \hline
 $ BP1 $ &  $\bf 1844\ih 487 $    & $1668\ih 212 $   & $1594\ih 162 $ & $ \redcross $ &  $ \redcross$ & $1119\ih 50 $ \\ \hline
 $ BP2 $ &  $\bf 2156\ih 637  $    & $ 1881\ih 212  $  & $ 1781\ih 162 $ & $\redcross $ & $\redcross $  & $1269\ih 62 $   \\ \hline
 $ BP3 $ &  $ \bf 2244\ih 675 $    & $1931\ih 200   $  & $1831\ih 162  $ & $\redcross $ & $\redcross $ & $1319\ih 75 $   \\ \hline
\end{tabular}
\caption{{\small Pole position for the benchmark points in Table \ref{table: BPs} varying the $\mathcal{O}(p^4)$ parameter $\gamma$.
    The rest of the parameters are set to their SM values. Values in boldface indicate broad resonances that do not satisfy $\Gamma<M/4$.}}\label{table: differences_coupled_gamma}
\end{table}\myspace
From the previous table, we can see that the appearance of a nonzero $\gamma$  makes the profile of the zero narrower in such a way
that we can even talk of a Breit-Wigner resonance with $\Gamma<M/4$. The values in boldface for $\gamma=0$ do not satisfy
this condition. In the case of an extreme value of $\gamma$, a value we would not expect for naturalness reasons, we recover
the single-channel approximation and, the coupled channel formalism is not necessary anymore. This is shown in the
last column for a value of $\gamma=10^{-2}$, where very narrow resonances appear.\myspace
\begin{table}[h!]
\begin{tabular}{|c|c|c|c|c|c|}
\hline
    $M_S\ih \Gamma_S$ & $\delta=0$  & $\delta=0.5\cdot 10^{-4}$ & $\delta=1\cdot 10^{-4} $ & $\delta=-0.5\cdot 10^{-4}$ & $\delta=-1\cdot 10^{-4}$  \\ \hline
 $ BP1 $ &  $\bf 1844\ih 487 $    & $ 1744\ih 362$   & $1669\ih 300   $   & $\bf 1994\ih 1100  $ & $ \doublepole $    \\ \hline
 $ BP2 $ &  $\bf 2156\ih 637 $    & $1981\ih 387   $  & $1869\ih 300  $   & $\bf 2644\ih \Gamma  $ & $\redcross $    \\ \hline
 $ BP3 $ &  $\bf 2244\ih 675$    & $2031\ih 400 $  & $1906\ih 287   $   & $\redcross $ & $\redcross $      \\ \hline
\end{tabular}
\caption{{\small Pole position for the benchmark points in Table \ref{table: BPs} varying the $\mathcal{O}(p^4)$ parameter $\delta$.
    The rest of the parameters are set to their SM values. Values in boldface indicate broad resonances that do not satisfy $\Gamma<M/4$.}}\label{table: differences_coupled_delta}
\end{table}\myspace
\begin{table}[h!]
\begin{tabular}{|c|c|c|c|c|c|}
\hline
    $M_S\ih \Gamma_S$ & $\eta=0$  & $\eta=0.5\cdot 10^{-4}$ & $\eta=1\cdot 10^{-4} $ & $\eta=-0.5\cdot 10^{-4}$ & $\eta=-1\cdot 10^{-4}$  \\ \hline
 $ BP1 $ &  $\bf 1844\ih 487$    & $1806\ih 437  $  & $1769\ih 387 $ & $\bf 1881\ih 575$ & $\bf 1931\ih 712$     \\ \hline
 $ BP2 $ &  $\bf 2156\ih 637$    & $2094\ih 512  $  & $2031\ih 437 $ & $\bf 2256\ih 887 $ & $\bf 2394\ih \Gamma$      \\ \hline
 $ BP3 $ &  $\bf 2244\ih 675 $    & $2156\ih 537  $  & $2094\ih 450 $ & $\bf 2356\ih 925 $ & $\bf 2544\ih \Gamma $      \\ \hline
\end{tabular}
\caption{{\small Pole position for the benchmark points in Table \ref{table: BPs} varying the $\mathcal{O}(p^4)$ parameter $\eta$.
    The rest of the parameters are set to their SM values. Values in boldface indicate broad resonances that do not satisfy $\Gamma<M/4$.}}\label{table: differences_coupled_eta}
\end{table}\myspace
The symbol $\redcross$ represents the absence of a zero in the determinant of the IAM matrix. We have also introduced
the symbol $\doublepole$ to indicate the situation where there are two poles in the unitarized amplitude but one is
unphysical following the phase-shift criteria; analytically, it corresponds to a pole in the
first Riemann sheet, which leads to a violation of causality with a negative width. Also, whenever our code is not able to
calculate the width over the profile of the "resonance" because it is too wide and the half maximum surpasses the HEFT validity,
we include the symbol ${\bf\Gamma}$, knowing that such a BP can never represent a physical resonance.\myspace
From the Tables \ref{table: differences_coupled_gamma}-\ref{table: differences_coupled_eta} above we can see a really different scenario from the one in the vector-isovector case. The location
of the pole changes $15\%-20\%$
when we use reasonable values of $\gamma$ and $\delta$ ($\sim 10^{-4}$) and softer variations of around $4\%-8\%$ for values
of $\eta$ of the same order. The lesson thus is clear: we cannot give a good description of the resonant scalar states
from $WW$ scattering without paying attention to the coupled channels.
In the first table, we have also included a big $\gamma$ value ($\sim 10^{-2}$) to make evident that the pole position in
that case is very similar to that obtained using the single-channel formalism neglecting the extra $I=0$ intermediate states.
In fact, for very non-natural values of $\gamma$ ($\sim 1$), the single-channel resonance is reproduced exactly.\myspace
The importance of the mixing parameters in determining the properties of the scalar resonances is now evident.\myspace
In the Tables \ref{table: differences_coupled_gamma}-\ref{table: differences_coupled_eta}, we have studied the effect in the resonance properties of the different couplings separately.
However, this may not be the general case since they are all independent and they are not strongly constrained (or even
constrained at all) by the experiment, especially the ones belonging to the Higgs sector, so they could all differ from zero.
Hence, it is not the individual effects but the simultaneous contribution of them all that we are interested in.
In Fig. \ref{fig: swept_gamma}, we show for the BPs in Table \ref{table: BPs} the space parameter in the $\delta - \eta$
plane where physical resonances with scalar masses heavier that $1.8$ TeV are allowed for different values of $\gamma$.\myspace 
No matter the value of $\gamma$ or the benchmark point selected, the presence of an unphysical pole appearing in the first Riemann sheet leads us to exclude the parameter space above the bands. This whole range of parameters cannot describe any physical extension of the SM. We also find that the greater the value of $\gamma$ is, the more restriction we find (there are more excluded space above the band), especially for BP1. \myspace
Below the bands, we find a nonresonant scenario: we do not find any zero in the determinant of the unitarized amplitude. 
\myspace
Because the above benchmark points correspond to relatively large masses, the amplitudes are to a large extent dominated by the NLO [i.e.
$\mathcal{O}(p^4)$] contributions. Those appearing in the $Wh$ mixed channel vanish when $\delta=\eta=0$, so the decoupling limit
results should be retrieved then.\myspace
The question of whether these resonances could be visible in the experiment requires a much more detailed study with Monte Carlo
techniques that is beyond the scope of this first study of scalar resonances. However, from the parton level processes studied
here, and by looking at the relative size of the residues of the corresponding poles in every channel, we can say
whether it is more likely to be a bound system of two $W'$s or a $hh$ composite state. Once the pole structure is factorized
from the unitarized amplitude, we are left with function which is a mixture of the other dynamical variables of the system; momentum structures
and couplings of the Lagrangian.\myspace
As an example, we show in Fig. \ref{fig: relative_residues} the amplitude of two unitarized amplitudes that show a
broad (left panel) and a narrow (right panel) resonances. In both cases they correspond to zeros of the determinant of the IAM amplitude.
We observe that the bigger the $\gamma$ parameter, the stronger the coupling to a $hh$ final state is, although the $WW$ channel is strongly
favored always. In any case, even if the dynamical resonances have a strong admixture of Higgs, they will be easier to spot in the $WW$ elastic
channel. This is a very clear prediction.\myspace
\myspace
To conclude this section, let us consider the case where some of the ${\cal O}(p^2)$ parameters differ from the SM limit. We shall
still keep $d_3=d_4=1$, but let us take $a=0.95$ and $b=0.805$. These values correspond to a minimal composite Higgs model living in the
subgroup $H=SO(5)/SO(4)$, which presents a symmetry in the Higgs function $\mathcal{F}(h)$ with the relation $b=2a^2-1$. The interested reader
may find in Ref. \citep{Dobado:2019fxe} a complete review for the different realizations of the HEFT, including this minimal extension.\myspace
In this case, and with the already mentioned expected maximum size of the $\mathcal{O}(p^4)$ of $\sim 10^{-3}$, we have not found any resonant
state that fulfills all the requirements of this study, even though some scenarios present resonant profiles that are too wide. All the physical
resonances for this choice of $a$ and $b$ appear for values of $a_{4,5}$ of order $10^{-2}$.\myspace
Not much can be concluded in this case.
\section{Bounds on $d_3$ and $d_4$ from resonances}
In this section, we will take the ${\cal O}(p^2)$ couplings $a,b$ to be equal to their SM value $a=b=1$ and explore how the resonance scene
depends on the triple and quartic Higgs couplings.
\subsection{$\mathbf{d_3}$}
The issue of determining the triple Higgs self-coupling is of utmost importance because it would help us to explore the properties
of the Higgs potential, crucial to understanding the nature of the Higgs boson itself. However, such a measurement is quite
involved at the LCH because it relies on the ability of the experiment to find a double Higgs final state (through its decay products)
coming from the fusion of two radiated (off-shell) electroweak gauge bosons or, alternatively, from top pairs. Up to now, not enough
statistics have been collected from the experiment, which translates into a very wide range in the experimental bound for this
coupling: $-3.3 < d_3 < 8.5$. The upper limit of this interval would make the interaction of $\mathcal{O}(1)$ since the BSM
self-interaction is described by $\lambda_3=d_3\lambda_{SM}$ with $\lambda_{SM}\sim 0.13$.\myspace
The fact that this coupling $d_3$ enters now at tree level in the calculation of the $I=0$ processes $Wh$ and $hh$ makes the resonant
scalar states in the spectrum of $WW$ scattering more sensitive to it and, hence, a good approach to the problem of investigating
the Higgs potential.\myspace
We start by analyzing the effect of this coupling separately, when the rest of the chiral parameters are set to their SM values, and
for the benchmark points in Table \ref{table: BPs}. The results are gathered in Table \ref{table: differences_coupled_d3}. \myspace
\begin{table}[h!]
\begin{tabular}{|c|c|c|c|c|c|c|}
\hline
    $M_S\ih \Gamma_S$  & $d_3=0.5 $ & $d_3=1$ & $d_3=2$ & $d_3=3 $ & $d_3=4 $ & $d_3=5 $  \\ \hline
    $ BP1 $     & $ \bf 2006\ih \Gamma $ & $\bf 1884\ih 487$ & $ 1681\ih 187 $  & $\doublerow{994\ih25}{1756\ih 65}$  & $\doublerow{1044\ih 38}{2069\ih 26}$ & $\doublerow{993 \ih 23}{2444\ih 25}$ \\ \hline
    $ BP2 $     & $ \bf 2369\ih \Gamma$ & $\bf 2156\ih 637 $ &$1906\ih 237 $  & $ \doublerow{1119\ih 27}{1869\ih 75} $ & $\doublerow{1219\ih 37}{2094\ih 31}$ & $ \doublerow{1181\ih 21}{2444\ih 25}  $  \\ \hline
 $ BP3 $ & $ \bf 2468\ih \Gamma$ & $\bf 2244\ih 675 $ & $1969\ih 250 $  & $\doublerow{1131\ih 19}{1894\ih 75}$ & $\doublerow{1269\ih 37}{2094\ih 20}$ & $\doublerow{1231\ih 23}{2444\ih 25}$  \\ \hline
\end{tabular}
\caption{{\small Values of the pole position of the benchmark points in Table \ref{table: BPs} changing $d_3$.
    The rest of the parameters are set to their SM values. The cells with two complex numbers indicate the pole position of the two physical Breit-Wigner poles in the denominator of the unitarized amplitude.}}\label{table: differences_coupled_d3}
\end{table}\myspace
We find that for $d_3\gtrsim 2.5$ a second pole clearly appears (notation pole1 over pole2) in the low-energy region
around $\sim 1$ TeV and it is also physical because it is found in the second Riemann sheet of the complex $s$ plane.
However, one of the physical poles is located at energy scales much lower than our preestablished bound of $1.8$ TeV, so,
in principle, the corresponding set of parameters should be discarded. The results are shown in Table \ref{table: differences_coupled_d3}. In fact, there are already hints of this first resonance at $d_3= 1.7$. \myspace
Of course, the possibility of a light scalar resonance ($\lesssim 1.8$ TeV) being very weakly coupled to $WW$ channel
and, hence, viable but hard to detect yet due to limited statistics remains a logical possibility to be further studied.
However, if we discard such possibility, the bound on $d_3$ becomes very stringent.\myspace
We have checked that the inclusion of a natural value of $\gamma$, does not alter the fact that one of the states is too light,
making the restriction on $d_3$ not significantly modified as it can be seen in Table \ref{table: differences_coupled_d3_gamma}. In this table, we reproduce the same analysis that we have just presented but set the value $\gamma=0.5\cdot 10^{-4}$.\myspace
\begin{table}[h!]
\begin{tabular}{|c|c|c|c|c|c|c|}
\hline
    $M_S\ih \Gamma_S$ & $d_3=0.5 $ & $d_3=1$ & $d_3=2$ & $d_3=3 $ & $d_3=4 $ & $d_3=5 $  \\ \hline
    $ BP1 $  & $1769\ih 275 $  & $1668\ih 212 $ & $ 1544\ih 112 $  & $\doublerow{994\ih 23}{1569\ih 25}$ & $\doublerow{1044\ih 37}{1769\ih 34}$ & $\doublerow{994\ih 27}{1994\ih 54}$  \\ \hline
    $ BP2 $  & $1981\ih 262 $ & $1881\ih 212 $  & $1719\ih 125  $  & $\doublerow{1106\ih 27}{1656\ih 50}$ & $\doublerow{1219\ih 37}{1781\ih 34}$ & $\doublerow{1118\ih 26}{1994\ih 50}  $  \\ \hline
 $ BP3 $ & $ 2031\ih 250$ & $1931\ih 200 $ & $1769\ih 125  $  & $\doublerow{1131\ih 37}{1681\ih 38}$ & $\doublerow{1269\ih 37}{1781\ih 27}$ & $\doublerow{1231\ih 23}{1994 \ih 53}$  \\ \hline
\end{tabular}
\caption{{\small Values of the pole position of the benchmark points in Table \ref{table: BPs} with $\gamma=0.5\cdot 10^{-4}$ changing $d_3$.
    The rest of the parameters are set to their SM values. The cells with two complex numbers indicate the pole position of the two physical Breit-Wigner poles in the denominator of the unitarized amplitude.}}\label{table: differences_coupled_d3_gamma}
\end{table}\myspace
The next step is to check the impact of the crossed channels by varying $\eta$ and $\delta$ in the phenomenological constraint found.\myspace
By doing so, we have not found any resonant state fulfilling all the criteria that we have imposed. For the three selected benchmark points of
Table \ref{table: BPs}, the behavior is quite similar and can be summed up in following three situations depending on the region in the $\eta-\delta$ plane: the firs scenario (1) with a single light resonance ($\sim 1$ TeV), another scenario (2) where two physical resonances appear but one is too light and a third new scenario (3) where a chain of three resonances emerge but the more massive one is classified as unphysical by the phase shift criteria. With all this, the bound $d_3\lesssim 2.5$ is not modified.

\subsection{$\mathbf{d_4}$}
The coupling $d_4$ parametrizes the strength of the self-interaction of four Higgses, and as it happens with $d_3$, it enters
now at the lowest order in chiral perturbation theory and contributes at tree level in the $hh$ process. From experiment, it is extremely poorly constrained because of
the difficulty of measuring the pointlike coupling of four Higgses. For this study and in the absence of any relevant experimental
bounds up to date, we will be considering values up to $d_4\lesssim 10$, which would make the interaction of order $\mathcal{O}(1)$.
Negative values of $d_4$ are to be excluded outright due to vacuum-stability reasons. \myspace
To start the analysis, we select the benchmark points from the tables above and see how the value of $d_4$
affects the properties of the poles. In particular, we focus on the case where $\gamma=0.5\cdot 10^{-4}$ which for all
scenarios allowed the presence of resonances satisfying $\Gamma<M/4$.
\begin{table}[h!]
\begin{tabular}{|c|c|c|c|c|c|c|c|}
\hline
    $M_S\ih \Gamma_S$ & $ d_4=0.5$ & $d_4=1$  & $d_4=2$ & $d_4=3 $ & $d_4=4 $ & $d_4=5 $ & $d_4=8 $ \\ \hline
    $ BP1 $ & $1794\ih 250 $ & $1668\ih 212  $    & $1494\ih 137 $  & $1381\ih 112 $ & $1306\ih 87  $ & $1256\ih 75 $ & $1169\ih 50 $ \\ \hline
    $ BP2 $ & $1981\ih 225 $ & $1881\ih 212  $    & $1719\ih 175 $  & $1606\ih 125  $ & $1531\ih 112  $ & $1481\ih 87 $ & $1381\ih 75 $ \\ \hline
 $ BP3 $ & $2031\ih 225 $ & $1931\ih 200  $    & $1781\ih 162 $  & $1669\ih 137  $ & $1594\ih 112  $ & $1544\ih 100 $ & $1444\ih 75$ \\ \hline
\end{tabular}
\caption{{\small Values of the pole position of the benchmark points in Table \ref{table: BPs} changing $d_4$
    with $\gamma=0.5\cdot 10^{-4}$. The rest of the parameters are set to their SM values.}}\label{table: differences_coupled_d4}
\end{table}\myspace
From Table \ref{table: differences_coupled_d4} we can say that, if all the rest of parameters are set to their SM values, we
could exclude values of $d_4\gtrsim 2$ for BP2 and BP3 and BP1 would be excluded since these parameters lead to
light resonances that should have already been seen. As always we assume (rightly or wrongly) that any scalar resonance above
1.8 TeV should have been observed. And as always, we also force the vector resonances,  if present, to be heavier than
that scale.\myspace
The question whether the crossed channel (with the parameters $\delta$ and $\eta$ leading at high energies) could affect this
result is depicted in the following graphs, where, for different values of $d_4$, we show the regions in the $\delta-\eta$ plane
where resonances $M_S>1.8$ TeV can appear.
\begin{figure}
\centering
\includegraphics[clip,width=7.0cm,height=6.5cm]{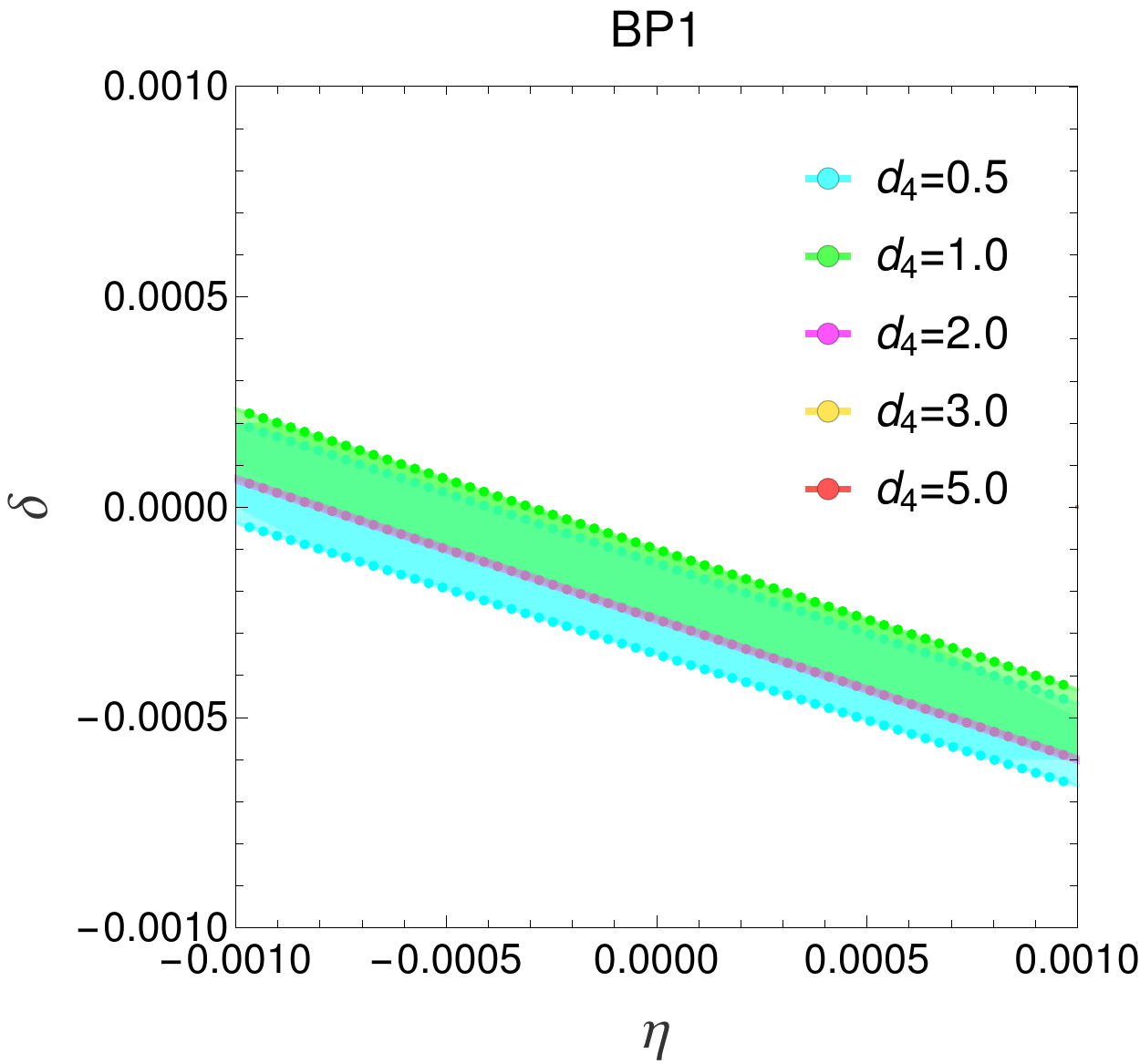} %
\includegraphics[clip,width=7.0cm,height=6.5cm]{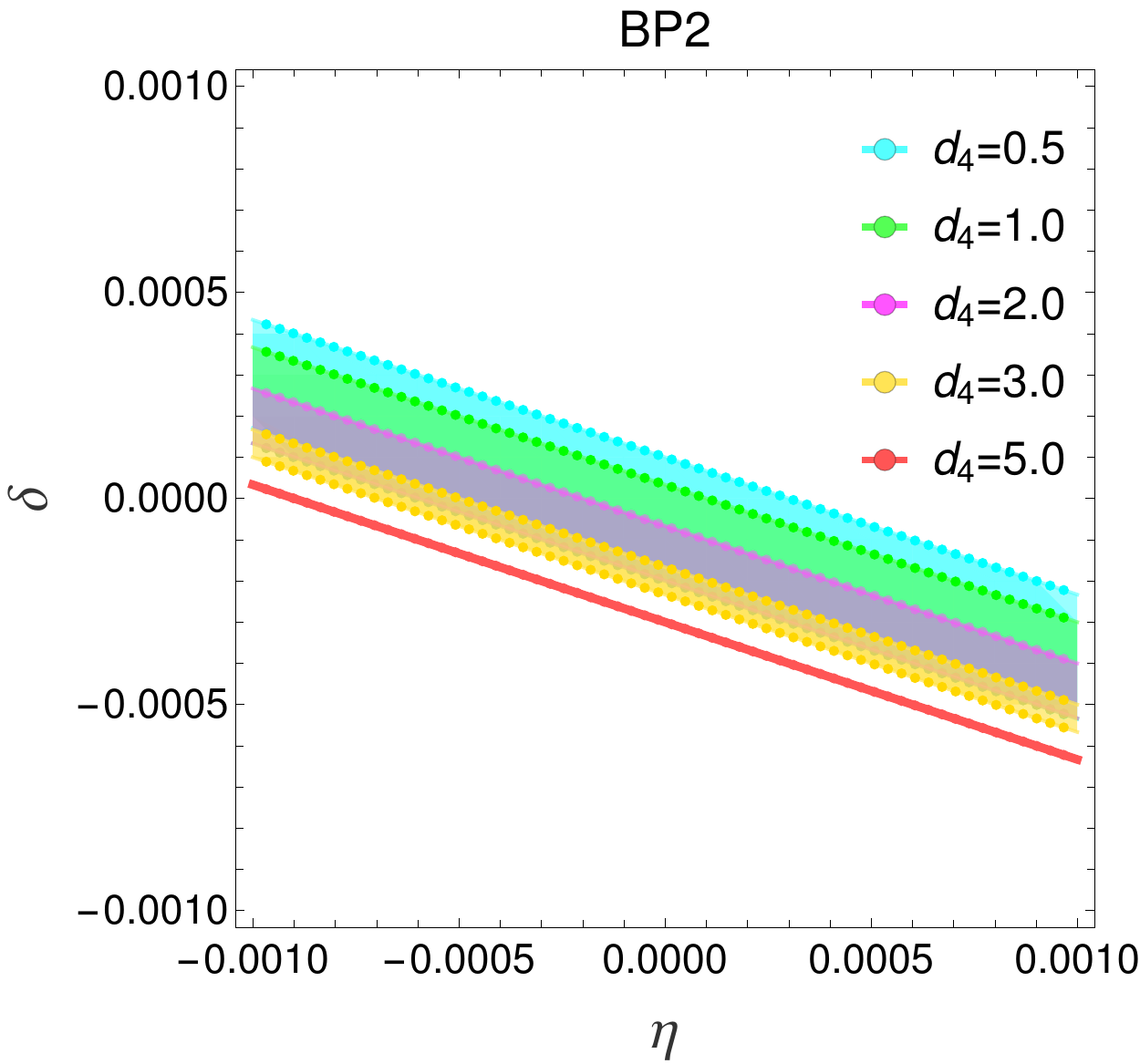} %
\includegraphics[clip,width=7.0cm,height=6.5cm]{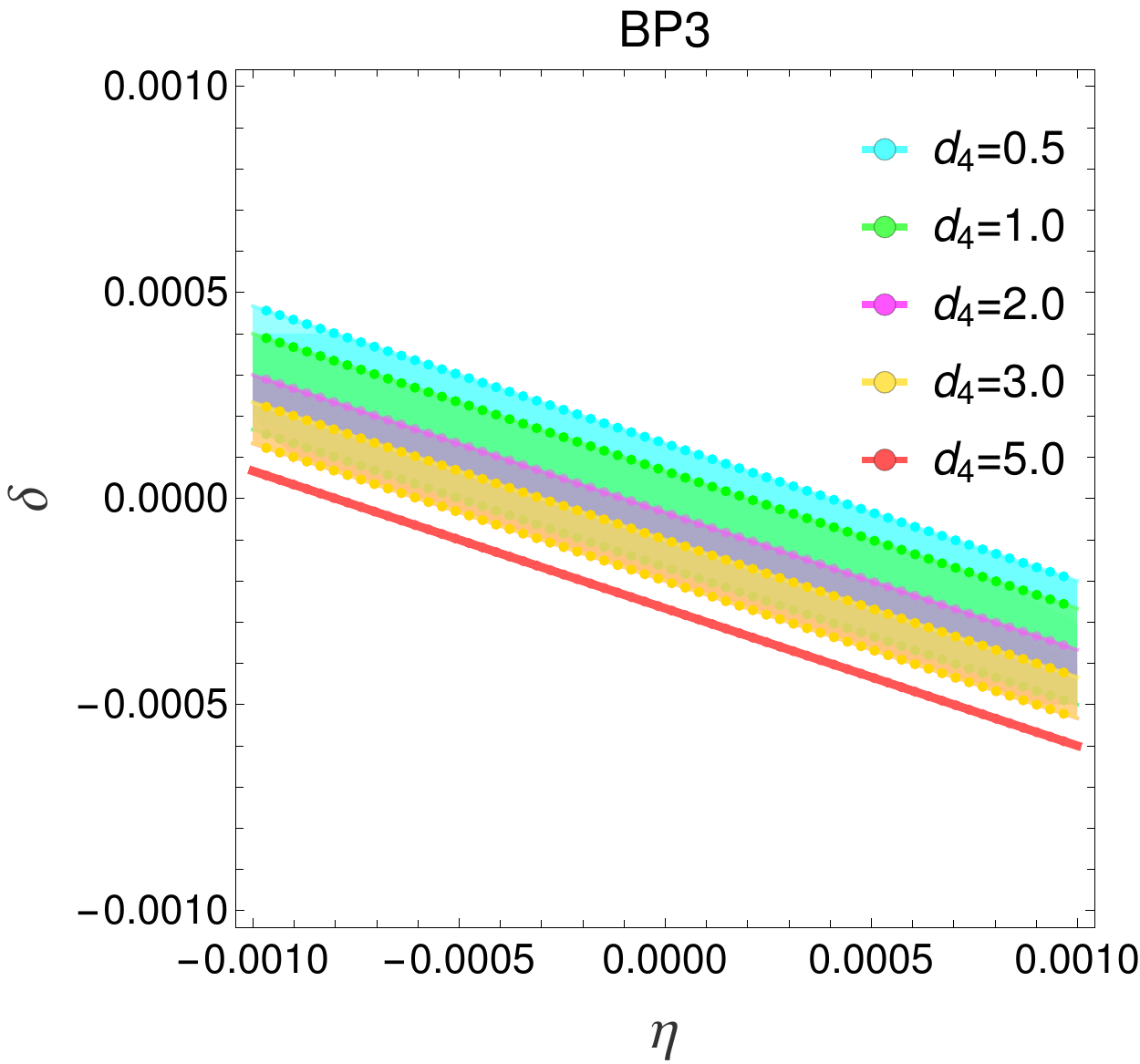} %
\caption{\small{Regions in the $\delta-\eta$ plane where physical resonances satisfying $M_S>1.8$ TeV appear for different
    values of $d_4$ and setting $\gamma=0.5\cdot 10^{-4}$ for specific values of $a_4$ and $a_5$ corresponding to (top left)
    BP1, (top right) BP2 and (bottom) BP3 in Table. \ref{table: BPs}. For all the values of $d_4\lesssim 6$ and for all the benchmark points, the region above the bands are excluded by the presence of a non-physical pole. Below the bands we find a non-resonant scenario.}}%
\label{fig: swept_BP2y3}
\end{figure}\myspace
In Fig. \ref{fig: swept_BP2y3}, we see how, in fact, some regions that were nonresonant, show resonances after activating the crossed-channels parameters from the values in the Table \ref{table: differences_coupled_d4}. The more we depart from the
SM value $d_4=1$, the more restriction we get. In fact, for $d_4=5$ and the values of Table \ref{table: differences_coupled_d4},
we only find resonances in the lines $\delta=-\frac{1}{3000}-\frac{\eta}{3}$ for BP2 and $\delta=-\frac{8}{3000}-\frac{\eta}{3}$
for BP3.\myspace
We do not find any physical resonant state with $M_S\gtrsim 1.8$ TeV and $d_4\gtrsim 6$.\myspace
In this case varying $d_4$, the same behavior that has been observed varying $\gamma$ (Fig. \ref{fig: swept_gamma}) is reproduced: above the color bands, we get excluded regions by the appearance of a second nonphyiscal pole (again using the phase-shift criteria), and below the bands, we get a nonresonant scenario with an absence of any zeros in the determinant of the unitarized amplitude.


\section{Conclusions}
Resonances are a characterist feature in a HEFT as
soon as one departs from the minimal Standard Model. Their role is to restore unitarity, and their properties, mass, and width and also their coupling to the various initial and final states are in close relation with the low-energy constants present in the HEFT.
Detecting one of such a resonance would undoubtedly signal the existence of additional microscopic degrees of physics but also point to particular regions
in the space of effective theories, suggesting fundamental physics of a certain kind.\myspace
Studying the properties of the possible resonances  provides precious information to experimentalist as to what type of signal is to
be expected in extensions of the Standard Model. It was seen in previous studies that the resonances appearing in $WW$ elastic scattering
are typically narrow and not very pronounced, indicating that, while the HEFT may be perturbatively nonunitary, it is, in a sense, close to
unitarity because the departures from the Standard Model are not large from a numerical point of view.
While this picture remains true in the vector resonance case, we have seen that when considering the $IJ=00$ case, where the formalism of
coupled channels is unavoidable when transverse gauge degrees of freedom are included, scalar resonances become substantially broader.\myspace
Assuming that no resonances exist below the scales that have already been experimentally probed, the next-to-leading $\mathcal {O}(p^4)$ coefficients
in the HEFT should be at most of order 10$^{-3}$ and probably of order 10$^{-4}$. In this work we have assumed that no resonance, vector or scalar,
exists below 1.8 TeV, and from that, we derive bounds on the HEFT couplings. But there is another way of restricting the HEFT; namely, if in the unitarization process one encounters acausal or unphysical resonances, the corresponding set of parameters in
the effective theory can be ruled out.\myspace
We assume three different conditions to characterize a resonance as physical: (1) $\Gamma < M/4$; (2) the set of parameters in the HEFT
must not produce vector or scalar resonances below $1.8$ TeV; and (3) all resonances (usually one, but sometimes two) must lie in the
second Riemann sheet. We also assume, as said,  that any resonance above 1.8 TeV should have been observed by now. \myspace
The space of parameters is fairly large, so the present study has to be understood only as a first exploration of this landscape that
surely merits an more systematic analysis. Let us nevertheless summarize the more relevant constraints.\myspace
First, we have verified  that not all ${\cal O}(p^4)$ coefficients are equally important. Those determining the appearance 
of resonances correspond to operators that survive in the nET limit. This, which is in agreement with previous
studies in the vector case, and it is quite useful as it tells us where to look for resonances in the vast space of HEFT.
Note that the inclusion  of transverse modes becomes relevant (unlike in the vector case)
in the scalar case. Second, besides $a_4$ and $a_5$, three new parameters appear at next-to-leading order.
Taking all the ${\cal O}(p^2)$ couplings to be identical to the SM, we have found that, for specific values of $a_4$ and $a_5$, the resonance
spectrum that could be observed is more restricted in the $\delta-\eta$ plane the lower the value of $\gamma$ is. For $\gamma=0$, resonances live in
a narrow band of values of $\delta$ and $\eta$, and the greater the value of $\gamma$, the broader the band is. The region above these narrow bands
can be excluded on causality grounds. So, this places a very strong restriction in parameter space. It is also seen almost immediately
that if a resonance is present in one channel it is present in all, but they always couple more strongly to $WW$ final states.\myspace
In this study, and making use of the arguments explained above, we have also set encouraging theoretical bounds on the self-interactions of the Higgs,
especially in the case of the triple self-coupling whose BSM deviations are parametrized via $d_3$ (in units of $\lambda_{SM}$). We have found
for this coupling that whenever it exceeds $d_3\sim 2.5$ a second very light pole appears, which we assume it would have already been detected
in the experiment. The emergence of this light pole becomes noticeable even before, from $d_3\sim 1.7$. The absence of such a resonance make
us exclude all the values above this threshold. This behavior is not significantly modified when considering nonzero values for the
parameters $\delta$, $\gamma$, and $\eta$, so from this study, we could set a bound $d_3\lesssim 2$, much more restrictive than current
experimental bounds (assuming, of course, that indeed no scalar resonance exists below 1.8 TeV). This is an important prediction; even if a
resonance is more likely to be observed in the $WW$ elastic channel, the Higgs self-coupling enters in the determination
of its properties via the coupled-channel formalism,  and it should not be dramatically different from its SM value.\myspace
For the case of the four-Higgs coupling, parametrized by $d_4$ (again in units of $\lambda_{SM}$), there are no experimental bounds in the literature to our knowledge. From this study of the single resonance spectrum, we have set an overall phenomenological bound $d_4\lesssim 6$ with regions
of the parameter space where it could be more restrictive, in particular for the point with both scalar and vector contribution (our BP1).\myspace
In conclusion, somewhat unexpectedly, the study of possible scalar resonances in $W W$ fusion places very interesting restrictions on the space of Higgs couplings,
a region that is hard to experimentally study. We have presented here some, we believe, relevant results, but certainly this line of research deserves further
more systematic studies.\myspace

\section*{Acknowledgments}

We thank Jose Ramon Pelaez, \`Angels Ramos, and Vincent Mathieu for useful discussions and clarifications.
We acknowledge financial support from the State Agency for Research of the Spanish Ministry of Science and Innovation
through the “Unit of Excellence Mar\'ia de Maeztu 2020-2023” award to the Institute of Cosmos Sciences (CEX2019-000918-M)
and  from PID2019-105614GB-C21, 2017-SGR-929 and 2021-SGR-249.

\appendix
\section*{Appendix A: Relevant counterterms}
\label{app:A}
In this Appendix we present the complete list of counterterms needed to absorb the one-loop divergences of the three relevant processes of this study. They have been obtained in the on-shell (OS) \citep{Grozin:2005yg} scheme and with the approximations mentioned in the previous sections of this piece of work; they are valid in the custodial limit, $g^{\prime}=0$, and within the Landau gauge ($\xi=0$) where the Goldstones are massless. In this scheme, the physical mass is placed in the pole of the renormalized denominator with residue 1. With all this, we just need to redefine two of the bare masses of the electroweak sector, $M_h$ and $M_W$, and no gauge parameter whatsoever due to its multiplicative renormalization.
\begin{equation}\label{eq: counter_masses_Zs}
\begin{split}
&\delta M_{h,div}^2=\frac{\Delta}{32 \pi^2 v^2}\left(3\left[6\left(2a^2+b\right)M_W^4-6a^2M_W^2M_h^2+\left(3d_3^2+d_4+a^2\right)M_h^4 \right]\right),\\
&\delta M_{W,div}^2=\frac{\Delta}{48 \pi^2 v^2}\left(M_W^2\left[3\left(b-a^2\right)M_h^2+\left(-69+10a^2\right)M_W^2\right]\right),\\
&\delta Z_{h,div}=\frac{\Delta}{16 \pi^2 v^2}\left(3 a^2\left(3 M_W^2-M_h^2 \right)\right),\\
&\delta Z_{\omega,div}=\frac{\Delta}{16 \pi^2 v^2}\left(\left(b-a^2\right)M_h^2+3 \left(a^2+2\right) M_W^2\right)
\end{split}
\end{equation}
where $\Delta=\frac{1}{\epsilon}+\log(4\pi)+\gamma_E$ represents the divergence.\myspace
In our setting, we let all the mass of the Higgs, the vev, and $\lambda_{SM}$ to get radiative corrections at NLO so the relation $M_H=2\lambda_{SM} v$ is jut valid at tree level. On the contrary, the relation $M_W=\frac{1}{2}gv$ is kept at all orders. This is why, in our case, it is meaningless to add a counterterm $\delta g$ since it is a derived quantity:
\begin{equation}
\frac{\delta g^2}{g^2}=\frac{\delta M_W^2}{M_W^2}+\frac{\delta v^2 }{v^2}.
\end{equation}
\myspace
In what concerns the chiral parameters, we just need to renormalize those couplings accompanying local operators of the scalar sector with no custodial-breaking pieces in them; these are $a,b,a_3,a_4,a_5,\gamma,\delta,\eta$, and $\zeta$:\\
\begin{equation}\label{eq: counter_coupling}
\begin{split}
&\delta v^2_{div}=\frac{\Delta}{16\pi^2}\left((b-a^2)M_h^2+3(a^2+2)M_W^2\right),  \quad \delta T_{div}=-\frac{\Delta}{32\pi^2 v}3\left(d_3M_h^4+6aM_W^4\right),\\
  &\delta a=\frac{\Delta}{32 \pi ^2 v^2}\left(6\,a \left(-2 a^2+b+1\right)M_W^2+(5a^3-a(2+3b)-3d_3(a^2-b))M_h^2\right), \\
&\delta b=\frac{\Delta}{32 \pi ^2 v^2}\left(6  \left(3 a^4-6 a^2 b+b (b+2)\right)M_W^2 \right. \\
&\qquad\left. -\left(21a^4-a^2(8+19b)+b(4+2b)+6ad_3(1+2b-3a^2)-3d_4(b-a^2)\right)M_h^2\right), \\
&\delta \lambda_{div}= \frac{\Delta}{64 \pi ^2 v^4}\left(\left(5a^2-2 b+3\left(d_3(3d_3-1)+d_4\right)\right)M_h^4 -12 \left(2 a^2+1\right) M_W^2 M_h^2\right.\\
&\qquad\left.+18  (a (2a-1)+b) M_W^4\right), \\
&\delta \lambda_3= \frac{\Delta}{64 \pi ^2 v^4}\left(36a b M_W^4+6 (3a^3-3ab-d_3(5a^2+1))M_W^2 M_h^2  \right. \\
&\qquad\left. +(-9a^3+3ab+d_3(10a^2-b)+9d_3d_4)M_h^4 \right), \\
&\delta \lambda_4=\frac{\Delta}{64\pi^2v^4}\left(36b^2M_W^4-12(a^2-b)(8a^2-2b-9ad_3)M_W^2M_h^2 \right.\\
&\qquad\left.+(96a^4+4b^2-d_3(114a^3-42ab)+9d_4^2+a^2(-64b+27d_3^2+12d_4))M_h^4 \right) ,\\
  &\delta a_3=-\frac{\Delta}{384\pi^2}\left(1-a^2\right),\quad \delta a_4=-\frac{\Delta}{192 \pi ^2}\left(1-a^2\right)^2, \\
 &\delta a_5=-\frac{\Delta}{768 \pi ^2}\left(5 a^4-2 a^2 (3b+2)+3 b^2+2\right),\\
  &\delta \gamma=-\frac{\Delta}{64\pi^2}3(b-a^2)^2, \quad \delta \delta = -\frac{\Delta}{192\pi^2} (b-a^2)(7a^2-b-6),
  \quad \delta \eta=-\frac{\Delta}{48\pi^2} (b-a^2)^2, \\
&\delta \zeta=\frac{\Delta}{96\pi^2}a(b-a^2)\,.
\end{split}
\end{equation}
All these counterterms have been proven to have the good SM behavior and to be consistent with the existing literature in the 
appropriate limits.


\begin{small}

\bibliographystyle{utphys.bst}
\bibliography{bibliography}

\end{small}

\end{document}